\documentclass[3p,authoryear,preprint,review,times]{elsarticle}

\usepackage[T1]{fontenc}
\usepackage{amssymb}
\usepackage{amsmath}          
\usepackage{array}
\usepackage{adjustbox}
\usepackage{gensymb}
\usepackage{caption}
\usepackage{multirow}
\usepackage{booktabs}
\usepackage{makecell}
\usepackage{subcaption}
\usepackage{bm}
\usepackage{natbib}
\usepackage{xurl} 
\usepackage[all]{nowidow}
\usepackage[hidelinks]{hyperref}

\usepackage[usenames, dvipsnames]{color}
\definecolor{myblue}{RGB}{0, 0, 205}
\definecolor{myred}{RGB}{205, 0, 0}
\definecolor{mypurple}{RGB}{155, 0, 155}
\definecolor{mygreen}{RGB}{0, 155, 0}
\definecolor{myturquoise}{RGB}{0, 155, 155}

\newcommand{\fig}[1]{Figure~\ref{#1}}
\newcommand{\tab}[1]{Table~\ref{#1}}
\newcommand{\eq}[1]{Equation~\ref{#1}}
\newcommand{\sect}[1]{Section~\ref{#1}}

\graphicspath{{figures/}}

\usepackage[switch]{lineno}

\journal{Fluid Dynamics Research}

\begin{document}

\begin{frontmatter}


\title{Propulsion of flapping foils undergoing in-plane clap-and-fling and deviation motions}

\author[1]{Antoine Papillon\fnref{label4}}
\author[1]{Mathieu Olivier\corref{cor1}\fnref{label5}}
\fntext[label4]{antoine.papillon.1@ulaval.ca}
\fntext[label5]{mathieu.olivier@gmc.ulaval.ca}

\cortext[cor1]{Corresponding author}
\ead{mathieu.olivier@gmc.ulaval.ca}

\affiliation[1]{
organization={Department of Mechanical Engineering},
addressline={Universit\'e Laval},
city={Quebec City},
state={QC},
postcode={G1V~0A6},
country={Canada}
}

\begin{abstract}
    This study examines the performance of two flapping flat-plate foils
    interacting with each other while generating thrust at a Reynolds number of
    800 through two-dimensional numerical simulations. These fluid dynamics
    simulations were conducted with a commercial computational fluid dynamics
    solver implementing a finite-volume method and an overset mesh capability.
    The foils performed a combined motion involving pitching, heaving, and
    deviation. Both foils exhibit similar movements, with one foil mirroring the
    other. The heaving and pitching motions occur at the same frequency but with
    a phase shift between them. The effects of varying the phase shift and the
    minimum spacing between the foils during motion were first explored. The
    study revealed that a maximum efficiency of 0.542 can be achieved by using
    two foils, representing an increase of 13.5\% relative to the optimal
    single-foil case. Then, the impacts of the deviation motion were
    investigated. The deviation motion was introduced with a frequency twice as
    fast as the other motions, and a phase shift relative to the heaving motion.
    The other parameters such as the minimum spacing between the foils, the
    heaving and pitching amplitudes, and the frequency were those of the best
    configuration without deviation. The numerical simulations demonstrated that
    deviation can increase efficiency further to a value of 0.560, a relative
    increase of 3.95\%.
\end{abstract}
\end{frontmatter}
%
%

\section{Introduction}
\label{Sec:intro}
The pursuit of leveraging useful aerodynamic forces from fluid flows has
captured the interest of researchers throughout the years. Numerous studies have
investigated flapping wings to either generate thrust or extract energy from
such flows \citep{tuncer_computational_1999, anderson_oscillating_1998,
miller_computational_2005, Wu2020}. This concept draws inspiration from the
efficient aerodynamics observed in various species of insects and small birds
that continuously flap their wings at specific frequencies to achieve flight.
This approach contrasts with intermittent flapping and gliding, a technique used
by most larger birds subject to larger Reynolds numbers. In some cases, an
intriguing phenomenon known as clap and fling is used. This mechanism was
initially reported by \cite{weis-fogh_quick_1973} who studied the flight of the
diminutive wasp \textit{Encarsia formosa}. They also demonstrated that this
clap-and-fling mechanism has been adopted by some birds and insects for flight.
It involves wings coming into proximity during flapping, resulting in enhanced
thrust or lift generation through their interaction.

\begin{figure}[!ht]
\centering
\includegraphics[width=0.6\columnwidth]{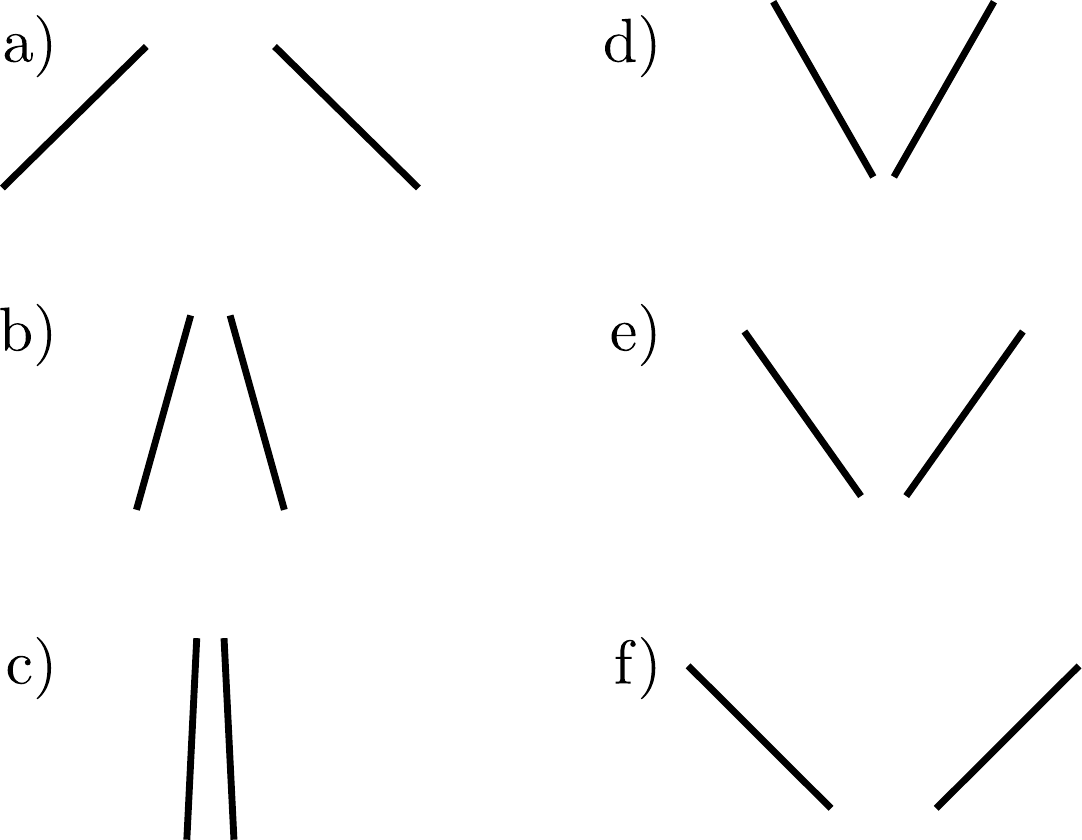} 
\caption{Schematic of wing sections during clap (a-c) and fling (d-f). The
motion occurs sequentially from (a) to (f) and the generated force is upward.}
\label{Fig_clap_fling_adapted}
\end{figure}

Conventional aerodynamic theories have often fallen short in explaining how
sufficient lift is generated at low Reynolds numbers, particularly for hovering.
Earlier studies identified that unsteady mechanisms play a crucial role in this
matter \citep{ansari_aerodynamic_2006, dickinson_wing_1999,
ellington_leading-edge_1996}. Among them, the clap-and-fling mechanism,
illustrated in \fig{Fig_clap_fling_adapted}, has been widely studied for its
efficient circulation generation around flapping wings through mutual
interaction. During the clap phase, the wings move toward each other and nearly
touch. This is followed by the fling phase, where the wings separate, inducing
circulation that rapidly reaches a steady state
\citep{lighthill_weis-fogh_1973}. Unlike traditional steady-state aerodynamics,
this mechanism bypasses the need for the Kutta condition to establish
circulation. The rapid reach of steady circulation allows the wings to maximize
lift during each oscillation cycle, explaining why several species evolved to
use this mechanism for flight at lower Reynolds numbers.

Building on these biological insights, computational and experimental studies
have advanced the understanding of flapping-wing aerodynamics.
\cite{platzer_aerodynamic_1993} analyzed thrust and lift in oscillating airfoils
using a panel code, revealing the impact of unsteady forces. Later work by
\cite{tuncer_computational_1999} introduced a Navier-Stokes solver to account
for flow separation and wake profiles, highlighting the role of attached flows
in achieving higher propulsive efficiencies. At higher plunging frequencies,
leading-edge vortices were observed to generate more thrust but at the cost of
reduced efficiency. These insights have inspired biomimetic designs, such as
developing flapping-wing vehicles with features mimicking species at comparable
Reynolds numbers \citep{jones_flapping-wing_2009}. Such designs leverage
unsteady aerodynamic advantages, including ground effects and wake ingestion, to
optimize thrust and efficiency \citep{platzer_flapping_2008}, underscoring the
continued relevance of biological flight mechanics to engineering innovation.

Meanwhile, numerous studies about flapping wings revealed the importance of
kinematic parameters on aerodynamic performances. In the case of harmonic wing
motions involving clap and fling, the phase shift between the heaving and
pitching movements has a significant impact since it controls the kinematics of
the foils when they reach proximity. For single harmonically flapping foils,
the optimal phase shift depends on the location of the pivot point on the
oscillating foil. \cite{anderson_oscillating_1998} have shown that this optimal
phase shift when the pivot point is at one-third of the chord length is
$75\degree$. They also pointed out that changing the pivot point location has
similar effects to modifying the phase shift in the case of small amplitude
motions. This is also supported by \cite{isogai_effects_1999}, who found that
the optimal efficiency occurs at a phase shift of $90\degree$ when the pivot
point is located at the middle of the chord length.

The impact of the interactions between two foils undergoing clap-and-fling
motion has been examined by some authors with two-dimensional numerical
simulations \citep{mao_flows_2003, miller_computational_2005, zhu_effect_2018}.
In the work by \cite{mao_flows_2003}, a hovering configuration with a Reynolds
number of 17 and a motion described by piecewise-continuous functions is used.
This configuration corresponds to the kinematics and physical scales of a wasp
(\textit{Encarsia Formosa}). They reported that reducing the minimum spacing
between the foils increases the lift coefficient. Yet, the moment coefficient,
which is directly related to the necessary power that needs to be injected to
accomplish the motion, increases significantly as the foils are closer
initially. \cite{mao_flows_2003} discussed that when the initial spacing varies
between the smallest values considered (10 to 20 percent of the chord length),
the lift coefficient barely diminishes while the moment coefficient becomes
much smaller. Hence, a compromise has to be made when setting the initial
spacing to maximize the lift while minimizing the power required to execute the
motion. They have also reported strong lift peaks during the clap phase, which
correspond to the moment when the trailing edges of the foils are close to each
other. They concluded that the clap-and-fling interaction can increase the lift
coefficient by a factor of up to 1.6 when compared to single foils. This
confirms, to some extent, that this mechanism can significantly improve lift.

Subsequently, \cite{miller_computational_2005} determined that a lower Reynolds
number could increase the lift of heaving and pitching foils. The spacing
between the two foils at the end of the fling movement was specified as 1/6 of
the chord length, which allows proximity effects to occur. Their simulations
were carried out for Reynolds numbers between 8 and 128. At lower Reynolds
numbers, lift improvements were reported during the translation and rotation
phases of the motion when compared to a single foil undergoing the same
movement. At higher Reynolds numbers, the propulsion improvements were
significantly reduced but kept a slight increase during the rotation part of
the motion. On the other hand, \cite{miller_computational_2005} also showed
that the necessary force to sustain the rotation motion increased at lower
Reynolds numbers. This implies that small insects likely need to provide
significant effort to rotate their wings. It was thus hypothesized that the use
of flexible wings could, however, decrease the negative impact of these drag
forces at lower Reynolds numbers.

Along the same lines, \cite{zhu_effect_2018} investigated the effects of
proximity and phase difference between the pitching and heaving of two
harmonically flapping foils. They considered a system moving with a Reynolds
number of $10^4$ rather than hovering. They reported a 77\% increase in
thrust and a 46\% increase in efficiency under different conditions. The best
thrust and efficiency were both obtained when the mean spacing between the
foils was 1.4 times the chord length, resulting in a minimum spacing between
the foils of about 30 percent of the chord length.

A study by \cite{wang_unsteady_2004} investigated unsteady effects and compared
2D computations with 3D experiments of different kinematic patterns for a
flapping wing in a hovering configuration. A dynamically scaled robotic fly was
used to obtain force and flow data experimentally, while a Navier-Stokes solver
was employed for the 2D simulations. The wing motion was harmonic and thus
followed a sinusoidal flapping and pitching motion. In the experimental case,
this motion was confined to an arc about the wing root, while in the 2D
computations, a simplification was considered such that the motion was along a
straight line. The main results of this study revealed that the force
coefficients have a weak dependence on stroke amplitude in the range considered,
which was 3-5 times the chord length. However, the forces were shown to be quite
sensitive to the phase between the flapping and pitching motions, which is
consistent with results from \cite{dickinson_wing_1999}. This can be explained
by the orientation of the wing at reversal which can optimally benefit from the
delayed stall mechanism. This increases the circulation around the wing and
consequently increases lift. However, \cite{wang_unsteady_2004} mentioned that
three-dimensional effects reported experimentally have an important role in this
mechanism.

Additionally, while numerous studies are based on harmonic motions, real fruit
flies benefiting from the clap-and-fling mechanism employ slightly more complex
kinematics. A comparison of several wing kinematic models based on 2D
Navier-Stokes simulations were performed by \cite{bos_influence_2008} to assess
the aerodynamic performances of a hovering insect. A harmonic model, a so-called
Robofly model based on the experimental device from \cite{dickinson_wing_1999},
a simplified fruit fly model, and a realistic fruit fly model were considered at
a Reynolds number of $110$. These four models were subjected to different stroke
patterns (heaving), angle of attack (pitching), and deviation, a motion of the
wing transverse to the main stroke motion. In the first two models, deviation is
not used while it is symmetric in the simplified fruit fly model and asymmetric
in the realistic fruit fly model.  Including deviation results in a
figure-of-eight pattern of the wing as viewed by an external observer. The study
shows that the mean drag at comparable lift for the harmonic and Robofly models
is substantially higher than in more realistic fruit fly models due to their
higher effective angle of attack. \cite{bos_influence_2008} also reported that
deviation could introduce an important velocity component perpendicular to the
stroke, which greatly affects the effective angle of attack of the wings.
Although the mean lift and drag were almost unaffected by deviation, their
amplitudes decreased. Therefore, the deviation motion levels the force
variations, which could be desirable for application to micro air vehicles
using such wing kinematics.

Additionally, wing kinematics of free-flying fruit flies were investigated
experimentally by \cite{fry_aerodynamics_2003} where three-dimensional infrared
high-speed videos were taken during the flight maneuvers. As recorded in this
study, the flight characteristics of real fruit flies include the deviation
motion responsible for the figure-of-eight pattern, the typical
wing stroke (heave in 2D), and the pitching motion. The data demonstrates
that each motion component is essentially harmonic except for the pitching
motion exhibiting more asymmetry. The most important takeaway of
this study is that the deviation motion is essentially in
phase with the stroke motion at a frequency twice as high. Furthermore, the
data presents these harmonic motions as angular displacements around the root
of the three-dimensional wing. In a 2D representation, these angular
displacements would thus have a corresponding simplified displacement. 

Deviation has also been studied through three-dimensional numerical simulations
carried out by \cite{jung_role_2024}. They considered a single flapping wing in
a hovering configuration with a similar motion to that of a mosquito, which
does not employ the clap-and-fling mechanism. Results demonstrated that for a
wing motion with a small sweeping amplitude (which is the wing displacement
around its root in 3D), the figure-of-eight pattern resulting from the deviation
significantly modified the effective angle of attack during a cycle of motion
and thus increased the mean lift by 30\%.

In this context, our study aims to explore further the clap-and-fling mechanism
of two flapping foils operating in a two-dimensional traveling configuration.
Contrary to hovering flight, this configuration implies an incoming flow
corresponding to the system velocity. The study focuses on the effects of the
phase difference between the heaving and the pitching motions as well as on the
minimum spacing between the foils.  Our study thus extends the work of
\cite{zhu_effect_2018} as we explore different parameters.  In particular, we
use a larger flapping amplitude and investigate configurations with
near-contact distances between the foils using an overset-mesh technique,
resulting in better thrust and efficiency. Furthermore, our study also
incorporates the deviation motion into the overall motion of the wings to
determine if beneficial effects can be obtained in cases with two wings in
proximity. To determine the optimal deviation for a two-dimensional
motion, we modified the phase shift between the heaving and deviation components
of the motion, as well as the amplitude of the deviation. By comparing the
results obtained through the inclusion of deviation to those previously
obtained without the latter, we investigate whether incorporating deviation can
also lead to higher levels of efficiency and thrust. These performances are
also compared against those of a single foil undergoing the baseline heaving
and pitching motions. This paper is divided as follows. Firstly,
\sect{Sec:methodology} presents the numerical methodology and the metrics that
characterize the performances of the airfoils. \sect{Sec:verification} presents
the verifications of the numerical simulations. Lastly, \sect{Sec:results}
reports and discusses the numerical results obtained with and without
deviation.

\section{Methodology}
\label{Sec:methodology}

\subsection{Problem definition}
\label{Sec:dynamics}

\begin{figure}[htb]
\centering
\includegraphics[width=0.45\columnwidth]{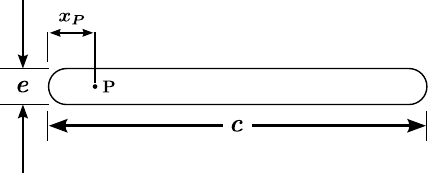} 
\caption{Geometry of a thin flat plate with rounded edges used as a foil.} 
\label{Fig_geometry_plate}
\end{figure}

This study employs two-dimensional numerical simulations of a pair of flapping
wings to investigate the combined effect of the deviation motion and the
clap-and-fling mechanism. By focusing on an in-plane motion, we can explore its
unique properties without considering 3D effects, which would require
substantial computational resources. Furthermore, considering that natural
flapping-wing flyers typically have thin wings and operate at low Reynolds
numbers, the impact of wing profile geometry is not significant in the proposed
study. Thus, the thin-plate geometry shown in \fig{Fig_geometry_plate} is used
as a foil where $c$ is the chord length, $e$ is the plate thickness, and $x_P$
represents the pivot point distance from the leading edge.

\begin{figure}[ht!]
\centering
\includegraphics[width=0.5\columnwidth]{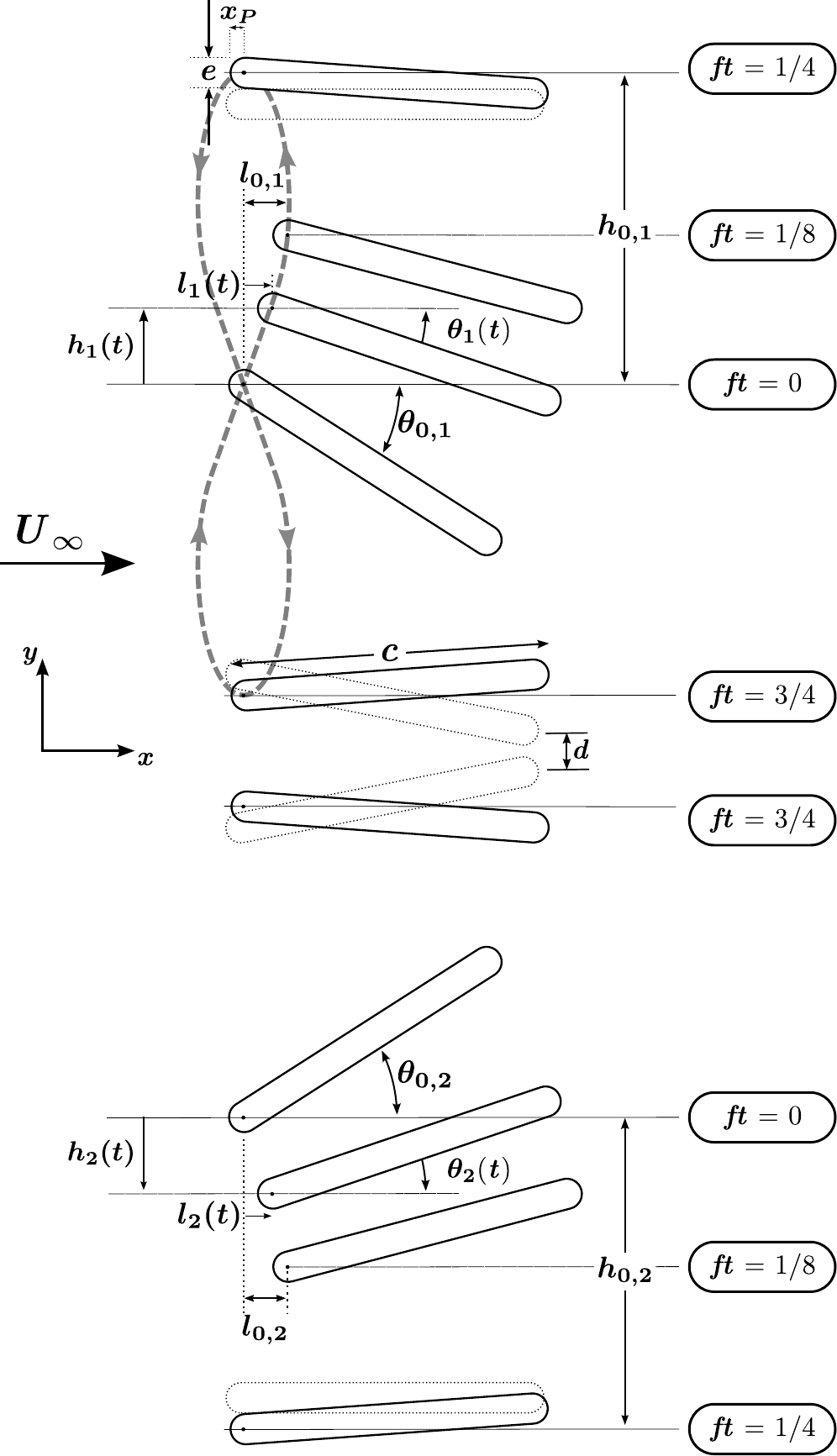} 
\caption{Motion of the two foils ($\phi=105 \degree$, $\psi=0 \degree$).}
\label{Motion_two_plates}
\end{figure}

The motion of the oscillating foils comprises three components: a vertical
translation (heaving motion), a rotation (pitching motion), and a horizontal
translation (deviation motion). Both foils undergo a symmetric motion as shown
in \fig{Motion_two_plates}, and the motion of the upper foil is described by:
\begin{equation}
h = h_0 \sin\left(2 \pi ft \right),
\label{eq_harmonic_1}
\end{equation}
\begin{equation}
\theta = \theta_0 \sin\left(2 \pi ft - \phi \right),
\label{eq_harmonic_2}
\end{equation}
\begin{equation}
l = l_0 \sin\left(4 \pi ft - \psi \right),
\label{eq_harmonic_3}
\end{equation}
where the heaving, pitching, and deviation amplitudes are denoted by $h_0$,
$\theta_0$, and $l_0$, respectively, and the motion frequency is represented by
$f$. Lastly, $\phi$ and $\psi$ are the phase differences between the heaving
motion and the pitching and deviation motions respectively. The complete motion
of the two flapping foils is illustrated in \fig{Motion_two_plates},
where the distance $d$ represents the minimum spacing attained between the
foils throughout the motion cycle. The incoming flow velocity, denoted as
$U_{\infty}$, corresponds to the baseline horizontal velocity of the system
without considering the effects of the deviation motion.

\subsection{Parametric Space}
\label{Sec:parametric_s}

The relevant reference scales used for this problem are the length scale $c$,
the velocity scale $U_{\infty}$, the time scale $1/f$, and the pressure scale
$\rho U_\infty^2$. With these scales, the following dimensionless parameters
that govern the problem are obtained:
\begin{equation}
\begin{split}
& Re=\frac{U_\infty \, c}{\nu}, \quad
f^*=\frac{f \, c}{U_\infty}, \quad
h^*=\frac{h_0}{c}, \quad
l^*=\frac{l_0}{c}, \quad \\
& e^*=\frac{e}{c}, \quad
x_P^*=\frac{x_P}{c}, \quad 
d^*=\frac{d}{c}, \quad
\theta_0, \quad
\phi, \quad
\psi.
\label{eq_parametric_space}
\end{split}
\end{equation}
These parameters are respectively the Reynolds number, the reduced frequency,
the normalized heaving amplitude, the normalized deviation amplitude, the
relative thickness of the airfoil, the normalized pivot point position, the
normalized minimum spacing between the foils, the pitching amplitude, the phase
shift between heaving and pitching, and the phase shift between heaving and
deviation. To focus on specific parameters, the following values are fixed
throughout the parametric study.
\begin{equation}
\begin{split}
& Re=800, \quad h^*= 1, \quad e^*=0.01, \quad x_P^*=0.005, \quad
\theta_0=30\degree, \quad f^*=0.15.
\label{eq_parametric_values}
\end{split}
\end{equation}
The Reynolds number is determined based on typical characteristics of flying
species presented in \tab{Tbl_flight_data}. The choices for normalized heaving
amplitude $h^*$ and normalized pivot point position $x_P^*$ are somewhat
arbitrary but enable the reproduction of a 2D motion resembling that of a
mid-span wing section. The pivot point position is essentially located at the
leading edge, which is inspired by birds’ wings whose motion is driven by the
forelimbs. The motion parameters $f^*$ and $\theta_0$ are selected to achieve
the best efficiency configuration when using a single wing with a phase
parameter $\phi = 90\degree$ (\cite{olivier_parametric_2016}). These parameter
choices leave the phase shift $\phi$ and the normalized minimum spacing between
the foils $d^*$ as the main parameters of interest when no deviation is used.
While the normalized distance $d^*$ certainly affects the clap-and-fling
mechanism, the phase shift $\phi$ influences the timing of the kinematics when
both wings come close together, which turns out to have a significant effect as
well. The two remaining key parameters in this study are the phase shift $\psi$
and the normalized deviation amplitude $l^*$. The normalized deviation amplitude
modulates the deviation motion relative to the heaving and pitching motions,
while the phase shift $\psi$ directly influences the shape of the
figure-of-eight pattern. The baseline phase shift considered is $\psi =
0\degree$, corresponding to the experimental flight data collected by
\cite{fry_aerodynamics_2003}.

\renewcommand{\arraystretch}{1.3}
\begin{table}[tb]
\caption{Flight characteristics of some flying species compiled from various
sources \citep{azuma_biokinetics_2006, tobalske_three-dimensional_2007,
weis-fogh_quick_1973}. Dimensionless numbers $f^{*}$ and $Re$ are computed with
the largest values of $U_{\infty}$.}
\begin{center}
\footnotesize
\label{Tbl_flight_data}
\begin{tabular}{@{}lcccccccccc@{}}
\toprule
Specie &
Mass (kg) &
$f$ (Hz) &
Span (m) &
$c$ (m) &
$U_{\infty}$ (m/s) &
$f^*$ &
$Re$ \\

\midrule

\makecell*[l]{Fruit fly \\ (\emph{Drosophila virilis})} &
$2 \times 10^{-6}$ &
240 &
0.006 &
0.0015 &
0 -- 0.65 &
0.55 &
$6.5 \times 10^{1}$ \\

\makecell*[l]{Mosquito \\ (\emph{Aedes nearcticus)}} &
$3.5 \times 10^{-6}$ &
320 &
0.0054 &
0.00067 &
0 -- 1.0 &
0.214 &
$4.4 \times 10^{1}$ \\

\makecell*[l]{Bee \\ (\emph{Bombus terrestris})} &
$8.8 \times 10^{-4}$ &
156 &
0.0346 &
0.0073 &
0 -- 2.41 &
0.473 &
$1.2 \times 10^{3}$ \\

\makecell*[l]{Hummingbird \\ (\emph{Selasphorus rufus})} &
$3.4 \times 10^{-3}$ &
41 &
0.109 &
0.012 &
0 -- 12 &
0.041 &
$9.6 \times 10^{3}$ \\

\bottomrule
\end{tabular}
\end{center}
\end{table}

\subsection{Performance metrics}
\label{Sec:perf_metrics}

The flapping-foil pair illustrated in \fig{Motion_two_plates} moves horizontally
with the deviation velocity $\dot{l}$. The foils's relative velocity with
respect to the flow is thus $U_{\infty}-\dot{l}$. When the foils translate
horizontally to the left, the effective velocity is thus greater than the
nominal velocity $U_{\infty}$ while the opposite occurs when the foils translate
to the right. Furthermore, each foil is subject to the aerodynamic force
$(F_{x},F_{y})$ and aerodynamic moment $M_{P}$. Considering the proposed
orientation for the $x$ and $y$ axes, the net thrust force which takes into
account the required force to perform the deviation motion is $-F_{x}$ while
$F_{y}$ and $M_{P}$ are the force and moment required to sustain the flapping
motion of each foil. The instantaneous thrust coefficient of each foil is thus
defined as follows:
\begin{equation}
    C_{T,i} = - \frac{F_{x,i}}{\frac{1}{2} \, \rho \, U_\infty^2 \, c}.
\label{Eq_Ct}
\end{equation}
Similarly, the power coefficient required to sustain the motion of foil $i$ is
defined as:
\begin{equation}
    C_{P,i} = -\frac{F_{y,i}\, \dot{h_i} \ + \ M_{P,i} \, \dot{\theta_i} \ + \ F_{x,i}\, \dot{l_i}}{\frac{1}{2} \, \rho \, U_\infty^3 \, c}.
\label{Eq_Cp}
\end{equation}
This equation is analogous to the one used by \cite{Stroke_deviation_2013}, and
it is derived under the hypothesis that the work required to sustain the wing
motion is provided by a system moving at $U_{\infty}$.  To account for the
unsteady nature of these force and power coefficients, a time-averaging
operation denoted by the operator $\overline{()}$ is performed once the flow
reaches a periodic behavior, typically after 3-5 cycles. This averaging is
carried out over the last 4 cycles of motion. The total average coefficients
considering both foils are expressed as follows:
\begin{equation}
    \overline{C_T} =  \frac{\overline{C_{T,1}} + \overline{C_{T,2}}}{2},
\label{Eq_Ct_glob}
\end{equation}
\begin{equation}
    \overline{C_P} =  \frac{\overline{C_{P,1}} + \overline{C_{P,2}}}{2}.
\label{Eq_Cp_glob}
\end{equation}

An additional important metric is the thrust efficiency of the flapping foils,
which represents the ratio between the thrust power $\left(F_x \,
U_{\infty}\right)$ and the power required to execute the heaving, pitching and
deviation motions. This efficiency is averaged over an integer number of
cycles, similar to the other coefficients, and is defined for one foil as:
\begin{equation}
    \eta_i = \frac{\overline{F_{x,i} \, U_{\infty}}}
    {\overline{F_{y,i}\, \dot{h_i} \ + \ M_{P,i} \, \dot{\theta_i} \ + \ F_{x,i}\, \dot{l_i}}}
    = \frac{\overline{C_T}}{\overline{C_P}}.
\label{Eq_eta}
\end{equation}

Since the motion of the two foils is symmetric, it is expected that $\eta_1 =
\eta_2$ most of the time. However, the resulting wake from the clap-and-fling
motion may not always be symmetrical. Therefore, another average efficiency
metric taking into account both foils is also considered:
\begin{equation}
    \eta =  \frac{\eta_1 + \eta_2}{2}.
\label{Eq_eta_glob}
\end{equation}

The last performance metric related to the deviation motion is the instantaneous
thrust coefficient standard deviation during one motion cycle. For foil $i$,
this metric is defined as: 
\begin{equation}
    \sigma_{C_{T,i}} =  \sqrt{\overline{\left(C_{T,i}-\overline{C_{T,i}}\right)^2}}.
\label{Eq_stand}
\end{equation}
The total propulsive standard deviation which considers both foils thus yields: 
\begin{equation}
    \sigma_{C_{T}} =  \frac{\sigma_{C_{T,1}} + \sigma_{C_{T,2}}}{2}.
\label{Eq_stand_glob}
\end{equation}
This metric correlates with how leveled the thrust coefficient is, which is
advisable in the case of micro air vehicles as discussed by
\cite{bos_influence_2008}. 

\subsection{Effective angle of attack}
\label{Sec:aero_metrics}

\begin{figure}[ht]
\centering
\includegraphics[width=0.5\columnwidth]{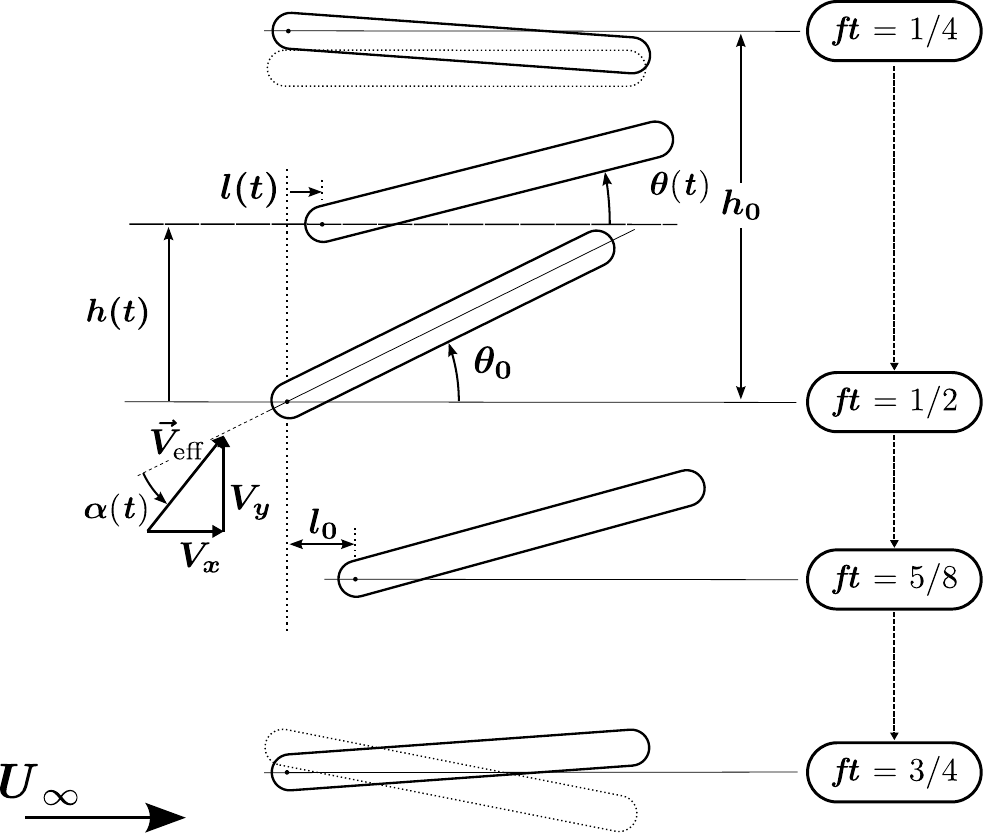} 
\caption{Imposed heaving $h(t)$, pitching $\theta(t)$ and deviation $l(t)$
motions for the upper foil ($\phi=105 \degree$, $\psi=0 \degree$). The
effective velocity $\vec{V_{\text{eff}}}$ and angle of attack
$\alpha$ are also shown.}
\label{fig_AOE}
\end{figure}

As discussed above, the flapping wing pair undergoes an effective horizontal
velocity which depends on the flow velocity $U_{\infty}$ and the deviation
motion. As a result, modifying the deviation parameters $\psi$ and $l^*$ as
defined in \sect{Sec:parametric_s} affects the flow behavior directly and
changes the effective angle of attack of the two foils. This necessarily impacts
the performance metrics described in \sect{Sec:perf_metrics} by affecting the
aerodynamic forces on the body.  Based on the motion imposed and upstream flow
conditions, the flapping foils pair undergoes an effective flow velocity
described as: 
\begin{gather}
\begin{gathered}
    \vec{V}_{\text{eff}}=V_x \, \hat{\textbf{\i}}+ V_y \, \hat{\textbf{\j}}, \\
    V_x=U_{\infty}-\dot{l}, \\
    V_y=-\dot{h},
\label{eq_effective_velocity}
\end{gathered}
\end{gather}
With this definition, the geometrical effective angle of attack $\alpha$ is
expressed at the pivot point position as:
\begin{equation}
    \alpha
    =\arctan\left(\frac{-\dot{h}}{U_{\infty} - \dot{l}}\right)-\theta.
\label{eq_AOE}
\end{equation}
On \fig{fig_AOE}, the effective angle of attack is thus shown to be the angle
between the chord line of the foil and the effective upstream velocity vector
$\vec{V}_{\text{eff}}$.

\subsection{Computational domain and numerical details}
\label{Sec:domain}

\begin{figure}[!ht]
\centering
\includegraphics[width=0.65\textwidth]{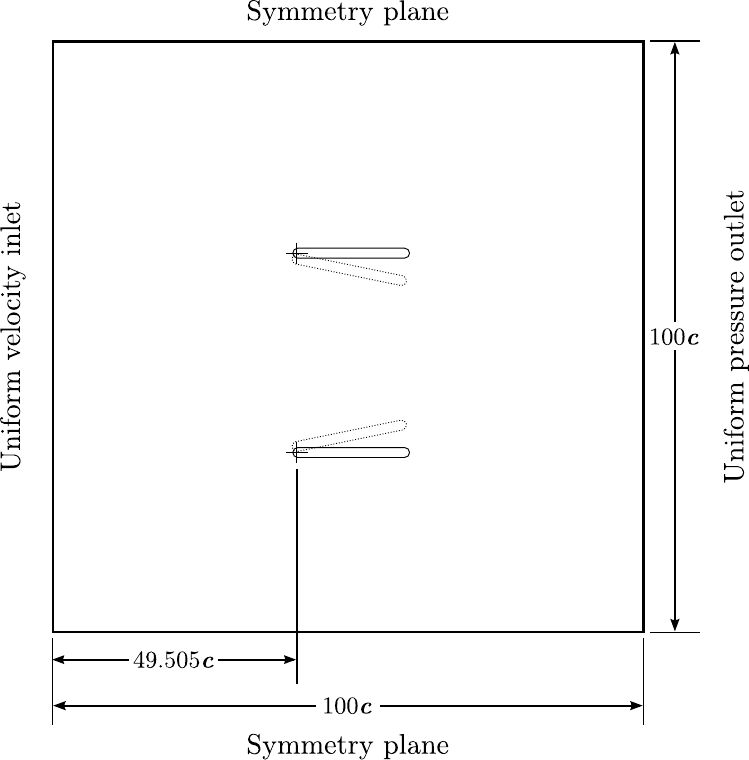} 
\caption{Computational domain of the simulation with two oscillating foils
(not to scale).}
\label{domain}
\end{figure}

\fig{domain} illustrates the computational domain used in this study, which
consists of a $100c$ $\times$ $100c$ box. The two foils are initially positioned
at a distance such that they reach the minimum spacing $d^*$.  The boundary
conditions applied are as follows: symmetry conditions are applied at the top
and bottom sides of the domain, a uniform velocity inlet condition is applied on
the left side, and a pressure outlet boundary condition is applied on the right
side with a uniform static pressure. The flow is governed by the incompressible
Navier-Stokes equations, and the numerical solutions are obtained with the
commercial solver STAR-CCM+.  The solver employs a second-order accurate
finite-volume method that supports unstructured polyhedral overset meshes. The
solution is advanced in time using a second-order backward difference scheme.

\begin{figure*}[!ht]
\centering
\begin{subfigure}[t]{.49\textwidth}
\centering
\includegraphics[width=\linewidth]{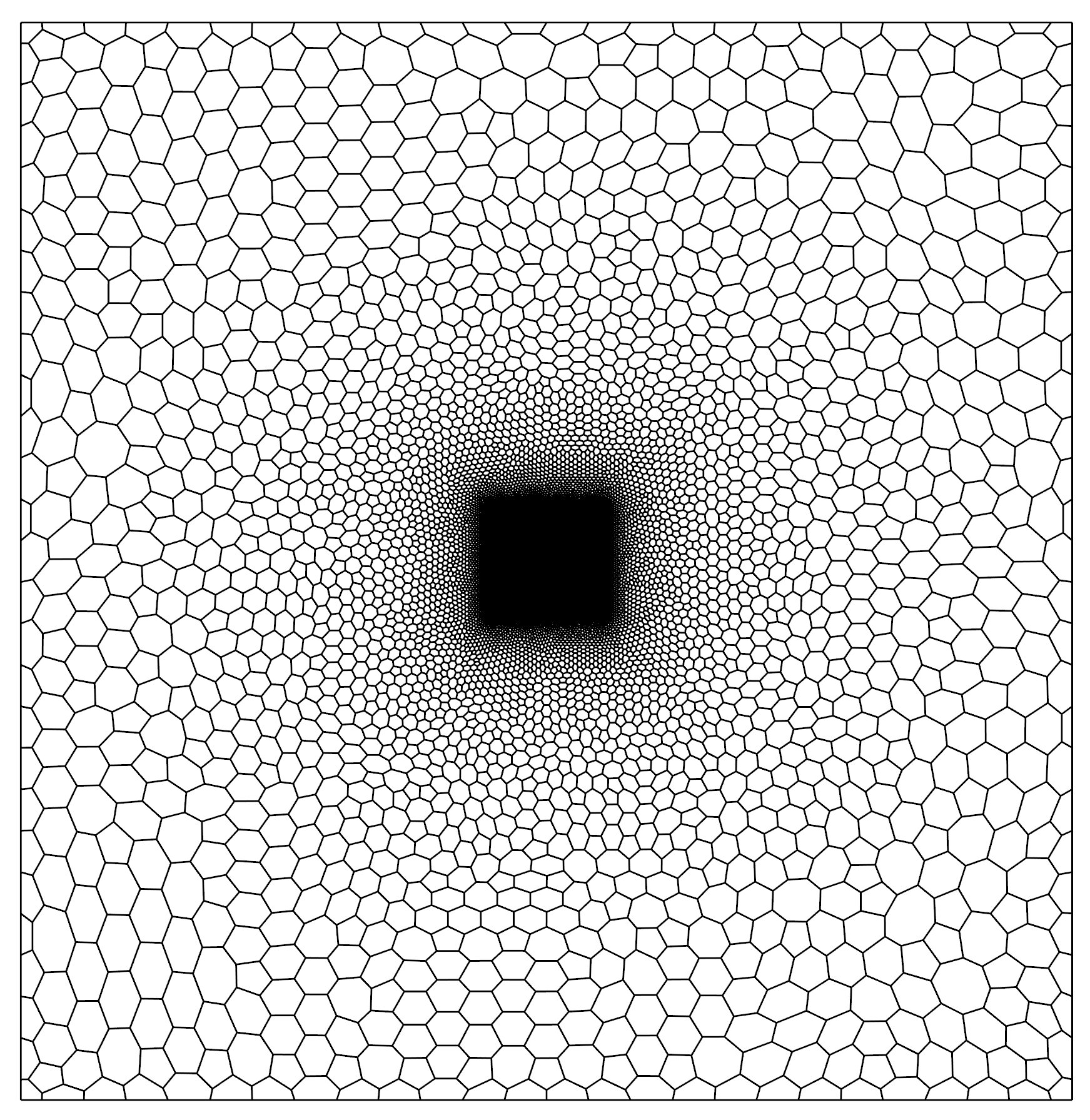} 
\caption{Background mesh used in the simulations.}
\label{mesh_amb}
\end{subfigure}
\hfill
\begin{subfigure}[t]{.49\textwidth}
\centering
\includegraphics[width=\linewidth]{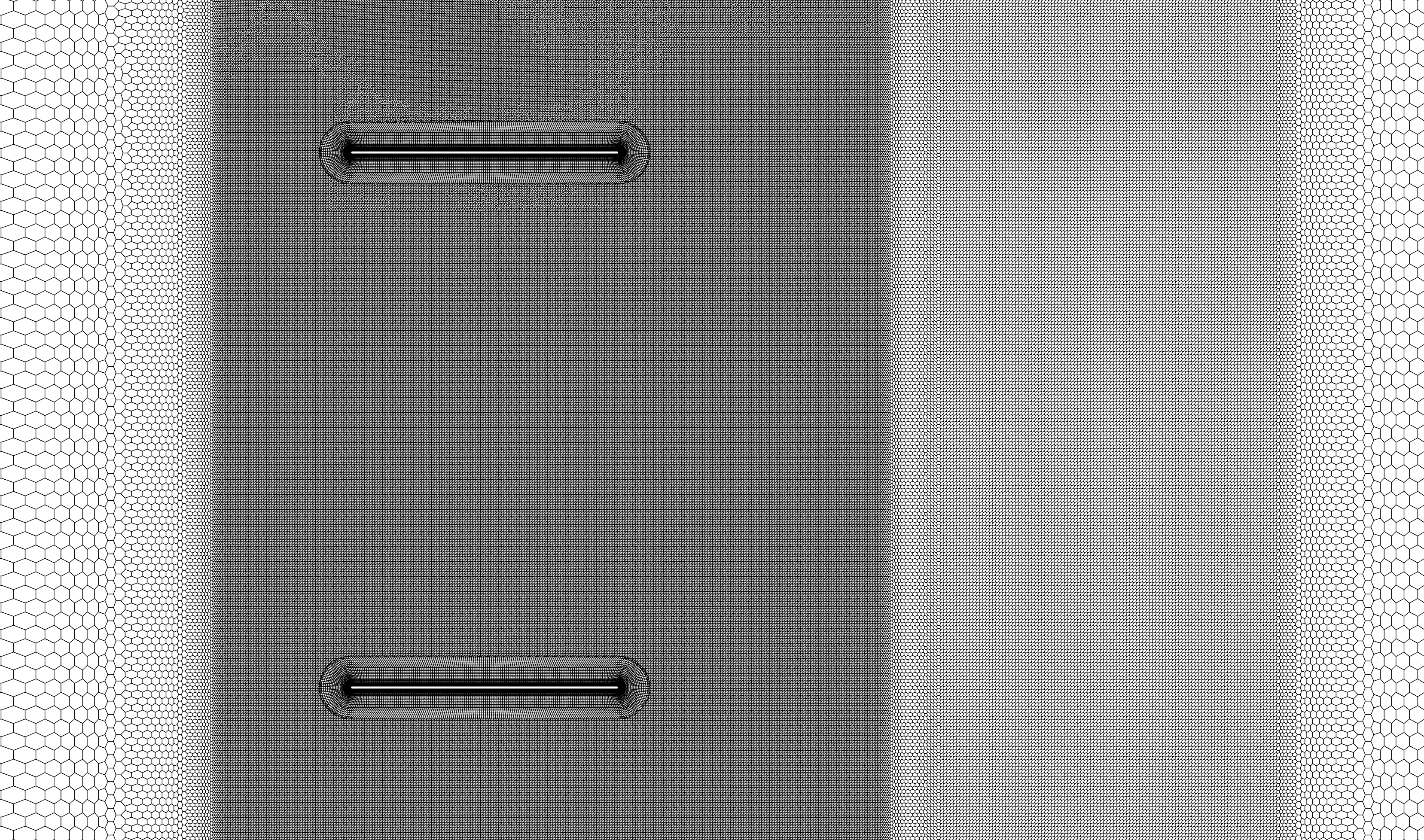} 
\caption{Mesh refinement near the two foils.}
\label{mesh_boite}
\end{subfigure}

\begin{subfigure}[t]{.75\textwidth}
\centering
\includegraphics[width=\linewidth]{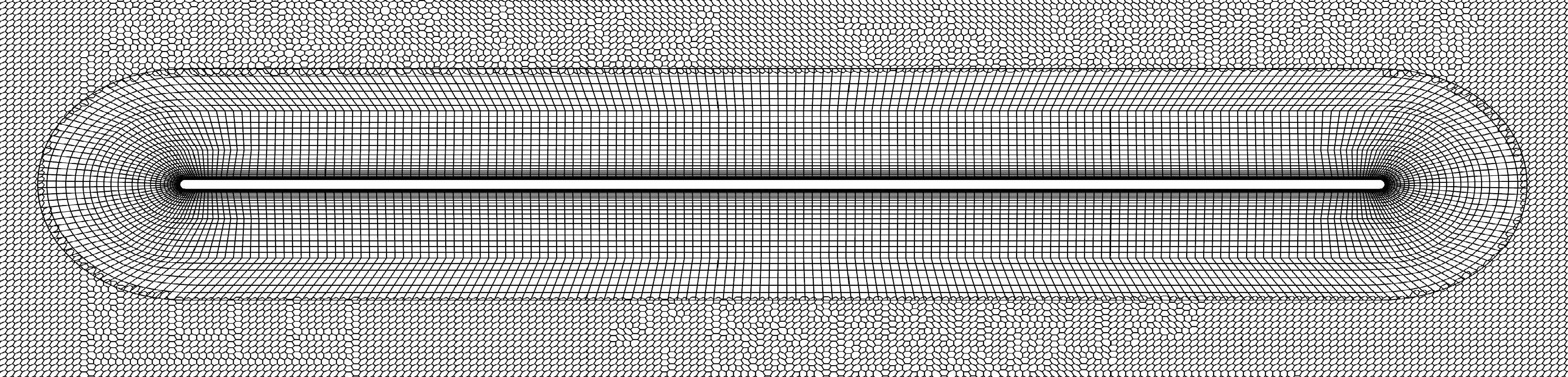} 
\caption{Overset mesh around a single foil.}
\label{mesh_plate_2D}
\end{subfigure}
\caption{Baseline mesh used in the simulations.}
\label{mesh_2D}
\end{figure*}

The mesh used for the simulations presented in this paper is displayed in
\fig{mesh_2D}. It consists of a polygonal background grid (\fig{mesh_amb}) with
a refined central region where the two foils are located. The refined region
includes two levels of refinement (\fig{mesh_boite}). The first level covers the
extent of the foils motion and has the same resolution as the overset meshes
around the foils (\fig{mesh_plate_2D}) to ensure accurate overset interpolation.
The second level, although slightly coarser, maintains a constant mesh
resolution to ensure proper resolution of the near wake. The overset mesh for
each foil is generated using the ICEM meshing software to have better control
over the meshing process. This allows the separation of each foil's mesh into
three sub-sections based on the distance from the wall. This separation protects
the smaller cells of the boundary layer from being ignored by the overset
interpolation when the different overset meshes come into proximity during the
motion. As a result, the interpolation between each sub-section uses cells of
similar size, enhancing the reliability of the results.


\section{Numerical verification}
\label{Sec:verification}

\subsection{Iterative convergence}
\label{Sec:convergence}

The flow solver uses the SIMPLEC iterative algorithm to handle the
pressure-velocity coupling. Therefore, appropriate stopping criteria must be met
at each time step to prevent significant iterative errors. The method we used in
this work is an asymptotic verification on three physical quantities: the
$x$-force, the $y$-force, and the moment computed on each foil (hence, there are
six stopping criteria for simulations involving two foils). The asymptotic
limits are normalized with five consecutive samples and are set to $10^{-4}$.
Thus, these criteria allow direct control of the impact of the iterations on the
physical quantities used for the performance evaluation.  Moreover, it was
verified that all equations' residuals followed a decreasing trend at each time
step of the simulations, thus complementing the asymptotic limit used as the
main stopping criterion. 

\subsection{Mesh and time step convergence}
\label{Sec:independence}

Three different meshes and time steps were employed to perform numerical
verification and assess the discretization error. These meshes included the
baseline mesh, which was used to generate the results, a coarser mesh, and a
finer mesh. The number of cells for each mesh is provided in
\tab{Tbl_verification_2D}. The time step values were adjusted to ensure a
consistent Courant number. The baseline dimensionless time step is $f \Delta t =
1/2000$.

\begin{figure}[htb]
\centering
\includegraphics[width=\linewidth]{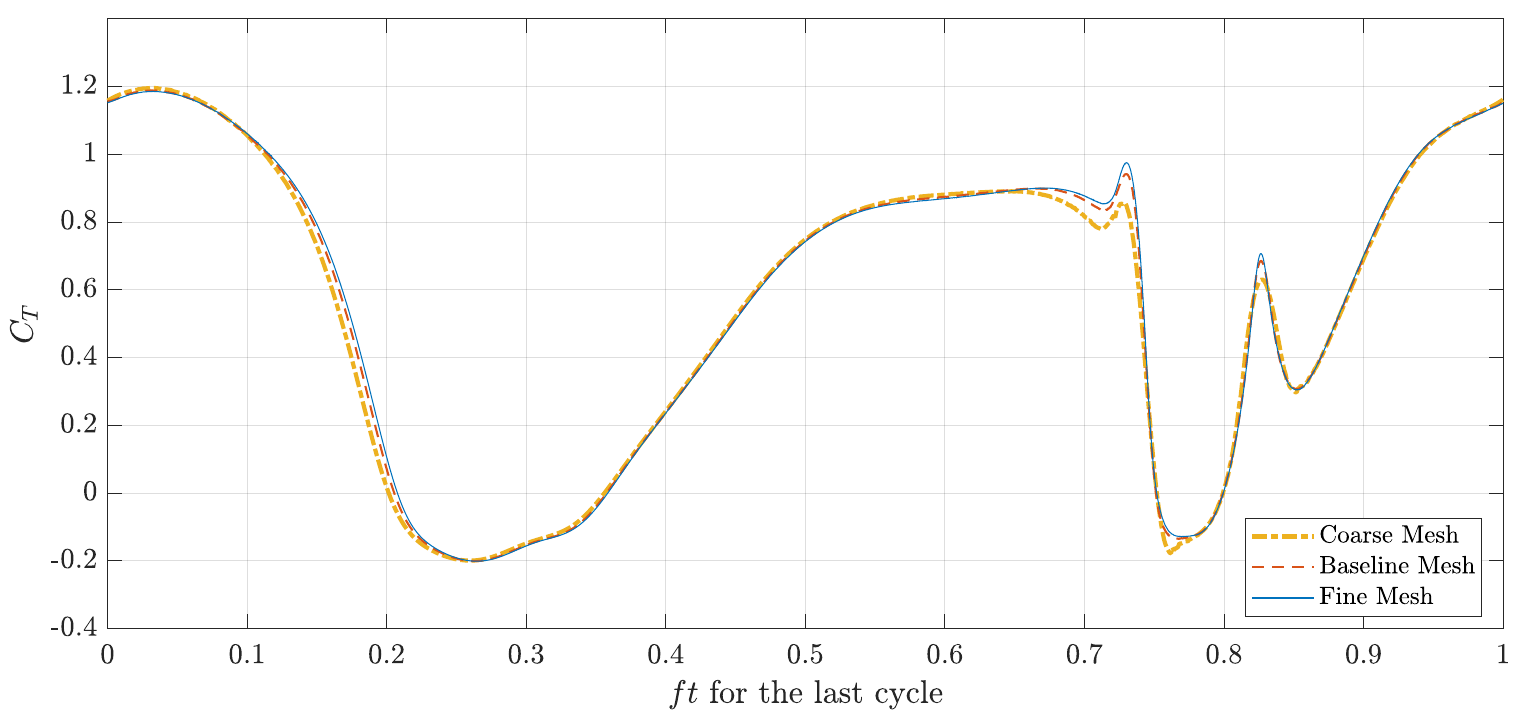} 
\caption{Convergence for the thrust coefficient for the last cycle of motion.}
\label{thrust_ev_glob_2D}
\end{figure}

\fig{thrust_ev_glob_2D} illustrates the variation of the thrust coefficient
during the last motion cycle for the three meshes and time steps. This
particular case corresponds to a pitching phase shift of $\phi=105\degree$, a
minimum spacing $d^*=0.005$, a deviation amplitude $l^*=0.100$ and a deviation
phase shift of $\psi=0\degree$. In this situation, the trailing edges of the
foils approach each other closely at the end of the clap phase.

Upon observing \fig{thrust_ev_glob_2D}, it is evident that the curves associated
with the different meshes and time steps exhibit strong similarity, indicating
good convergence. However, it is also apparent that the curve corresponding to
the coarser mesh displays larger discrepancies, particularly near the peaks and
troughs of thrust. This behavior is expected in a converged analysis and further
validates the suitability of the baseline mesh and time step.  The results from
\tab{Tbl_verification_2D} support these observations by comparing the
performance metrics introduced in \sect{Sec:perf_metrics}. They indicate that
simultaneous spatial and temporal refinement has minimal impact on the reported
physical parameters in the table, confirming the adequacy of the chosen baseline
mesh and time step.

\begin{table}[htb]
\caption{Discretization details and cycle-averaged thrust and power coefficients.}
\begin{center}
\footnotesize
\label{Tbl_verification_2D}
\begin{tabular}{lccccc}
\toprule

&
Total cell count &
\makecell{Overset zone \\ cell count} &
$\Delta t \cdot f$  &
$\overline{C_T}$ & $\overline{C_P}$ \\

\midrule
Coarse Mesh   & 180~199 & 4~140 & 1/1000 & 0.561 & 1.080 \\
Baseline Mesh & 672~549 & 11~776 & 1/2000 & 0.568 & 1.091 \\
Fine Mesh     & 2~680~535 & 39~432 & 1/4000 & 0.571 & 1.094 \\
\bottomrule
\end{tabular}
\end{center}
\end{table}

Further verifications were made to assess the accuracy of the overset mesh
interpolation, particularly during the time interval when the leading and
trailing edges of the foils come very close. In \fig{thrust_ev_glob_2D}, two
thrust peaks are observed around $ft=0.730$ and $ft=0.825$, coinciding with
substantial overlap between the overset meshes of each foil. This raises
concerns regarding the number of cells in the spacing between the two foils.
Indeed, a so-called hole-cutting procedure is executed by the solver during the
simulation when the overset meshes overlap.  This procedure activates and
deactivates mesh cells depending on the proximity of the foils and their
respective meshes, which can influence the mesh resolution and the quality of
the mesh-to-mesh interpolation in that region. To address this concern,
\tab{Tbl_verification_loc_2D} presents a verification wherein local averages for
the power and thrust coefficients are computed for the last cycle of each
simulation within the normalized time interval $ft \in [0.715,\ 0.850]$. This
interval corresponds to the two thrust peaks where the leading edges get close
to each other, followed by the trailing edges that reach even closer proximity
resulting from the subsequent fling of the foils. The variation in thrust and
power coefficients is sufficiently low to conclude that an adequate level of
convergence has been achieved, even when the foils are near and their respective
meshes overlap. Based on these verifications, it can be determined that the
baseline mesh with the associated normalized time step of $\Delta t \cdot f =
1/2000$ is appropriate for the parametric study.

\begin{table}[htb]
\caption{Verification of the discretization in the time interval
$ft\in [0.715,\ 0.850]$.}
\begin{center}
\footnotesize
\label{Tbl_verification_loc_2D}
\begin{tabular}{lccccc}
\toprule
&
Total cell count &
\makecell{Overset zone \\ cell count} &
$\Delta t \cdot f$  &
$\overline{C_T}\,^1$ & $\overline{C_P}\,^1$ \\

\midrule
Coarse Mesh   & 180~199 & 4~140 & 1/1000 & 0.276 & 0.545 \\
Baseline Mesh & 672~549 & 11~776 & 1/2000 & 0.294 & 0.565 \\
Fine Mesh     & 2~680~535 & 39~432 & 1/4000 & 0.302 & 0.574\\
\bottomrule
\multicolumn{6}{l}{\footnotesize $^1$ $\overline{C_T}$ and $\overline{C_P}$ values were averaged on the normalized time interval $[0.715,\ 0.850]$.} \\
\end{tabular}
\end{center}
\end{table}

\section{Results and discussion}
\label{Sec:results}

\subsection{Parametric Study Overview}
\label{Sec:parametric_study}
In this section, the effects of varying the four parameters of interest
presented in \sect{Sec:parametric_s} are investigated. This is done according to
a heuristic optimization methodology, thus by first varying the phase shift
between the heaving and pitching motions as well as the minimum spacing between
the airfoils without considering deviation. After identifying an optimal case by
using the constant parameters defined in \eq{eq_parametric_values} while varying
the phase shift between 60$\degree$ and 120$\degree$ and the minimum spacing
$d^*$ between 0.005 and 8, a second parametric investigation is performed by
including the deviation motion. The latter is performed by modifying the
deviation amplitude $l^*$ between 0 and 0.150, and the phase shift $\psi$
between -140$\degree$ and 40$\degree$. While the best case identified through
this heuristic optimization may not be the absolute optimum, it remains a highly
efficient configuration that yields valuable insight into the clap-and-fling
mechanism.

\subsection*{Cases without deviation}
\begin{figure}[!ht]
\centering
\includegraphics[width=0.98\linewidth]{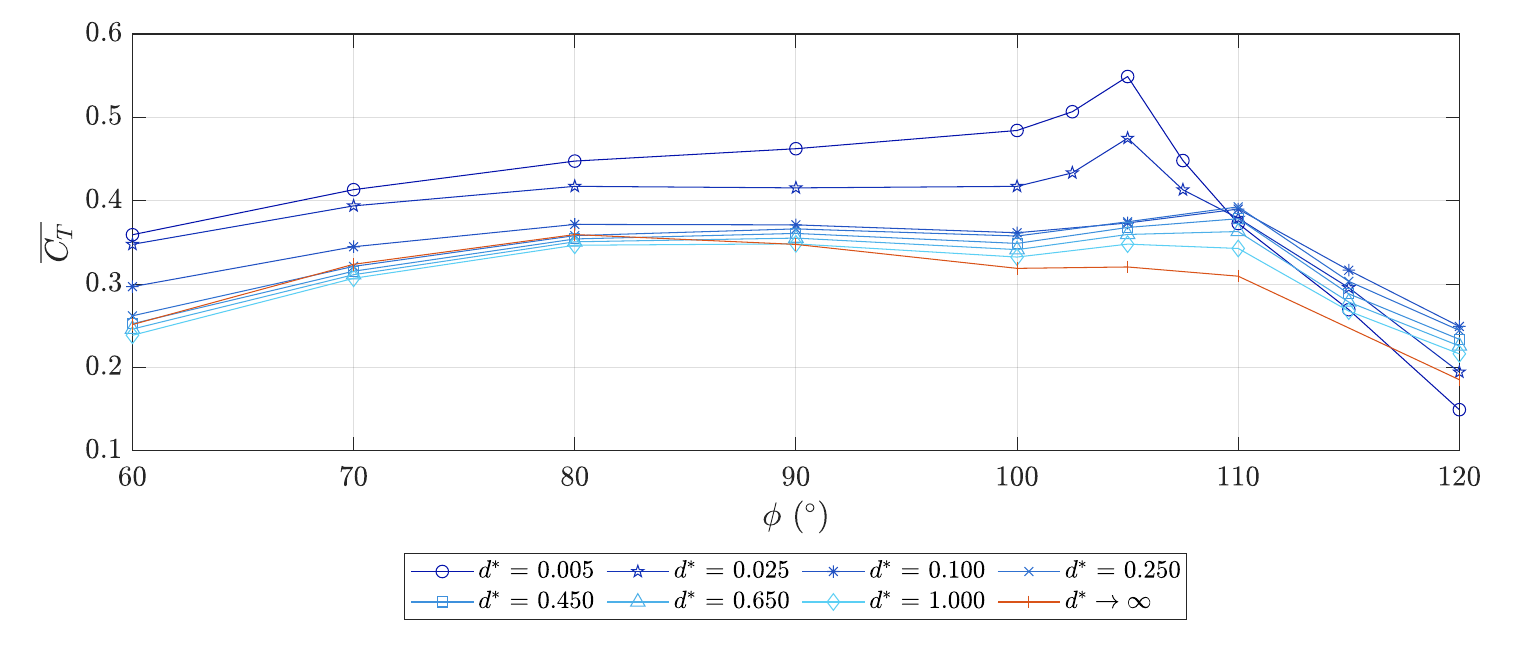} 
\caption{Averaged thrust coefficient $\overline{C_T}$ as a function of the phase
shift $\phi$ for different minimum spacings between the foils $d^*$ (no deviation).}
\label{CT_phase}
\end{figure}

\begin{figure}[!ht]
\centering
\includegraphics[width=0.98\linewidth]{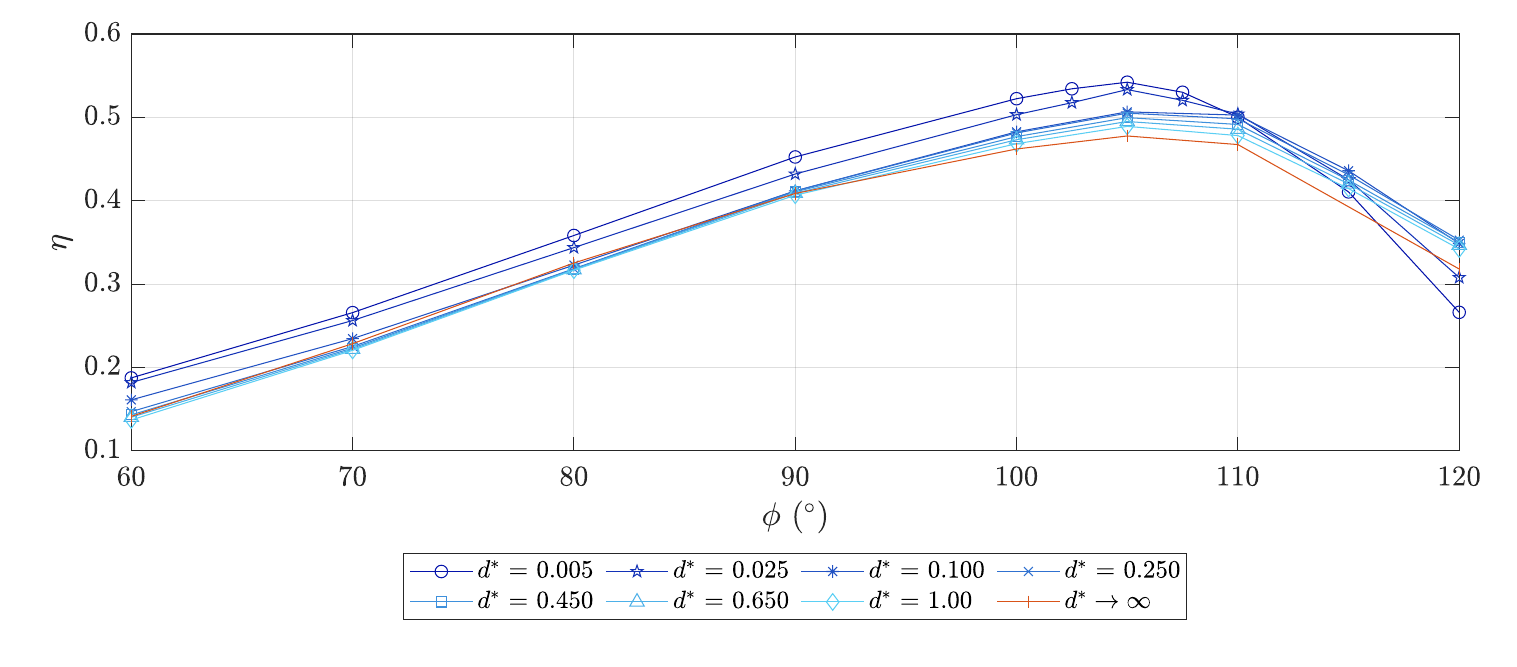} 
\caption{Efficiency $\eta$ as a function of the phase shift $\phi$ for different
minimum spacings between the foils $d^*$ (no deviation).}
\label{eta_phase}
\end{figure}

\begin{figure*}[!ht]
\centering
\begin{subfigure}[t]{.49\textwidth}
\centering
\includegraphics[width=\linewidth]{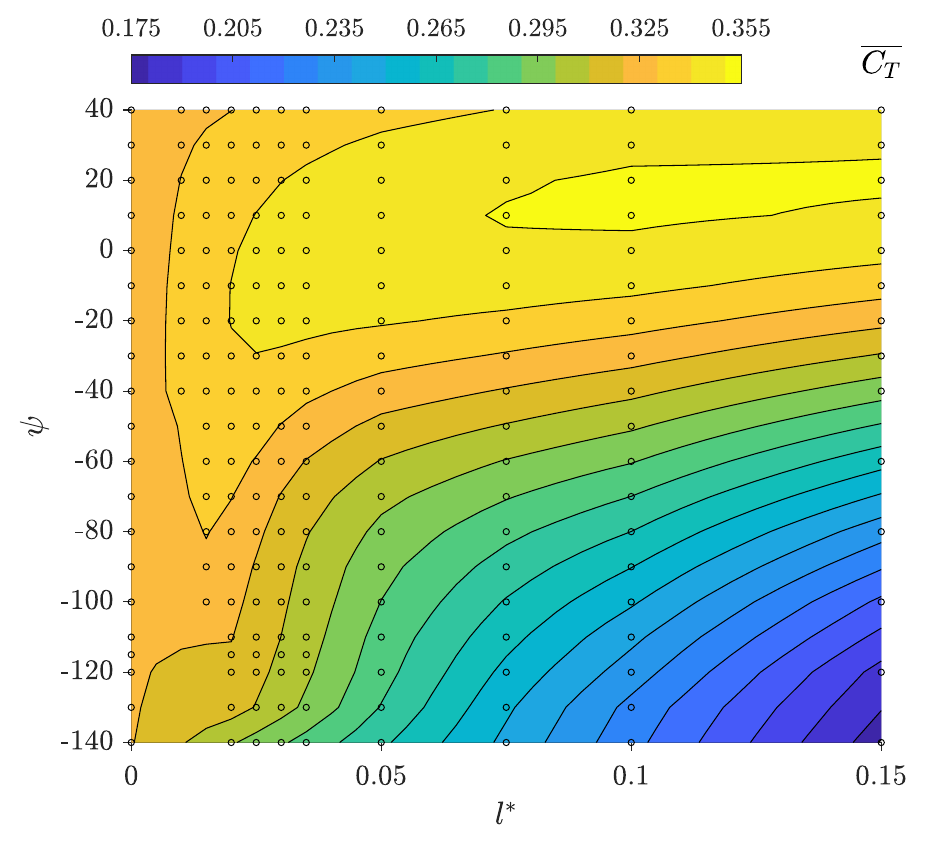}
\caption{Single foil.}
\label{CT_single_plate}
\end{subfigure}
\hfill
\begin{subfigure}[t]{.49\textwidth}
\centering
\includegraphics[width=\linewidth]{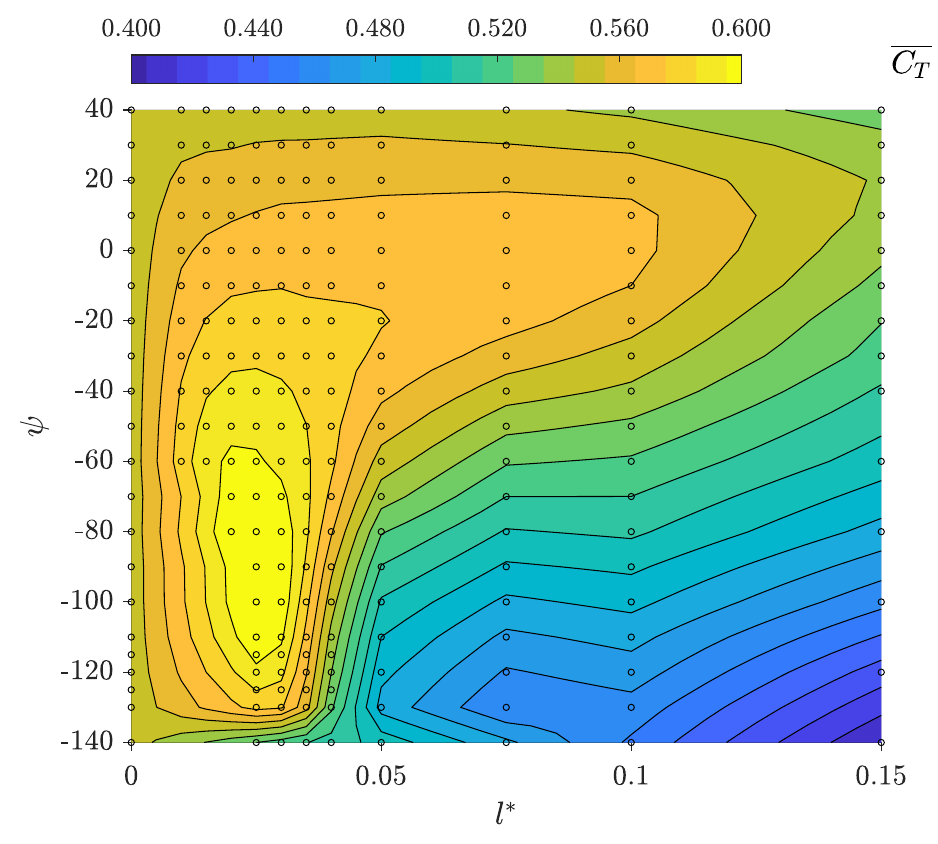}
\caption{Dual foils ($d^*=0.005$).}
\label{CT_dual_plate}
\end{subfigure}

\caption{Averaged thrust coefficient $\overline{C_T}$ as a function of the
deviation amplitude $l^*$ and phase shift between the deviation and heaving
motions $\psi$.}
\label{Colline_CT}
\end{figure*}

\begin{figure*}[!ht]
\centering
\begin{subfigure}[t]{.49\textwidth}
\centering
\includegraphics[width=\linewidth]{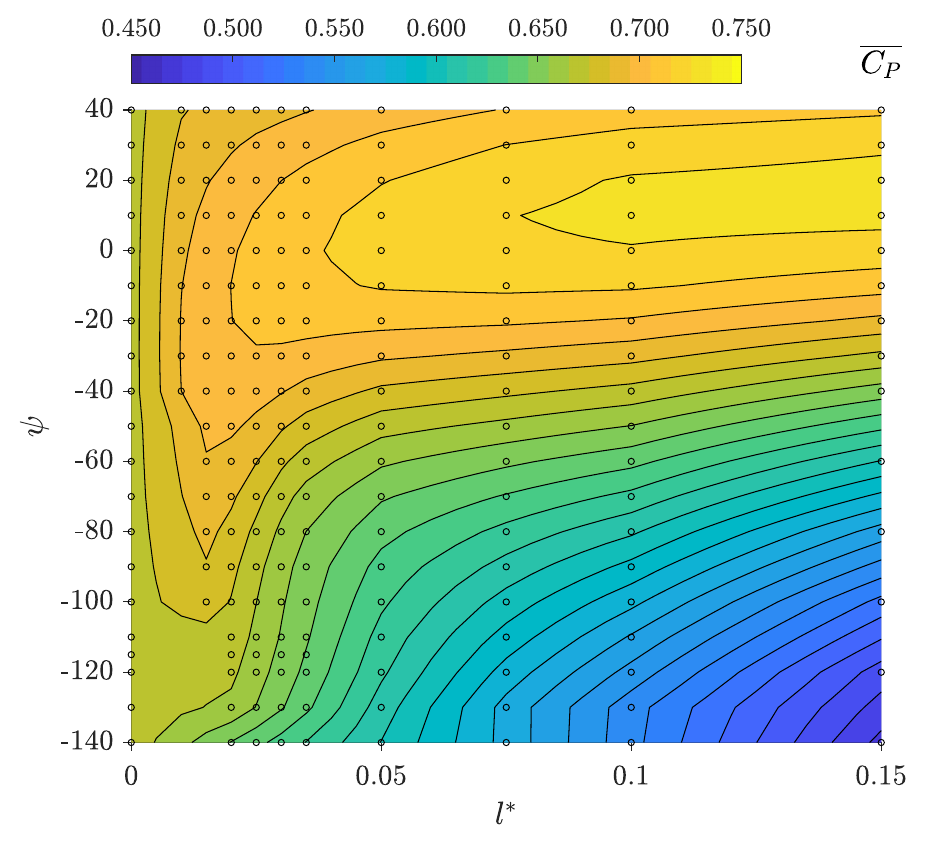}
\caption{Single foil.}
\label{CP_single_plate}
\end{subfigure}
\hfill
\begin{subfigure}[t]{.49\textwidth}
\centering
\includegraphics[width=\linewidth]{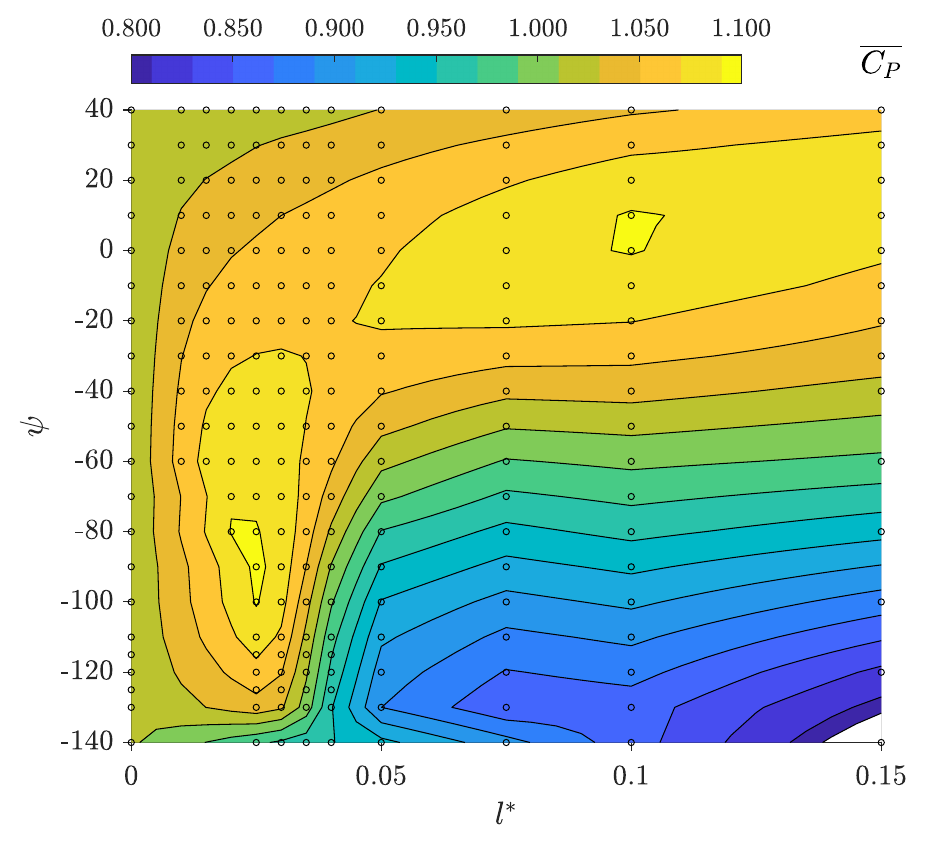}
\caption{Dual foils ($d^*=0.005$).}
\label{CP_dual_plate}
\end{subfigure}

\caption{Averaged power coefficient $\overline{C_P}$ as a function of the
deviation amplitude $l^*$ and phase shift between the deviation and heaving
motions $\psi$.}
\label{Colline_CP}
\end{figure*}

\begin{figure*}[!ht]
\centering
\begin{subfigure}[t]{.49\textwidth}
\centering
\includegraphics[width=\linewidth]{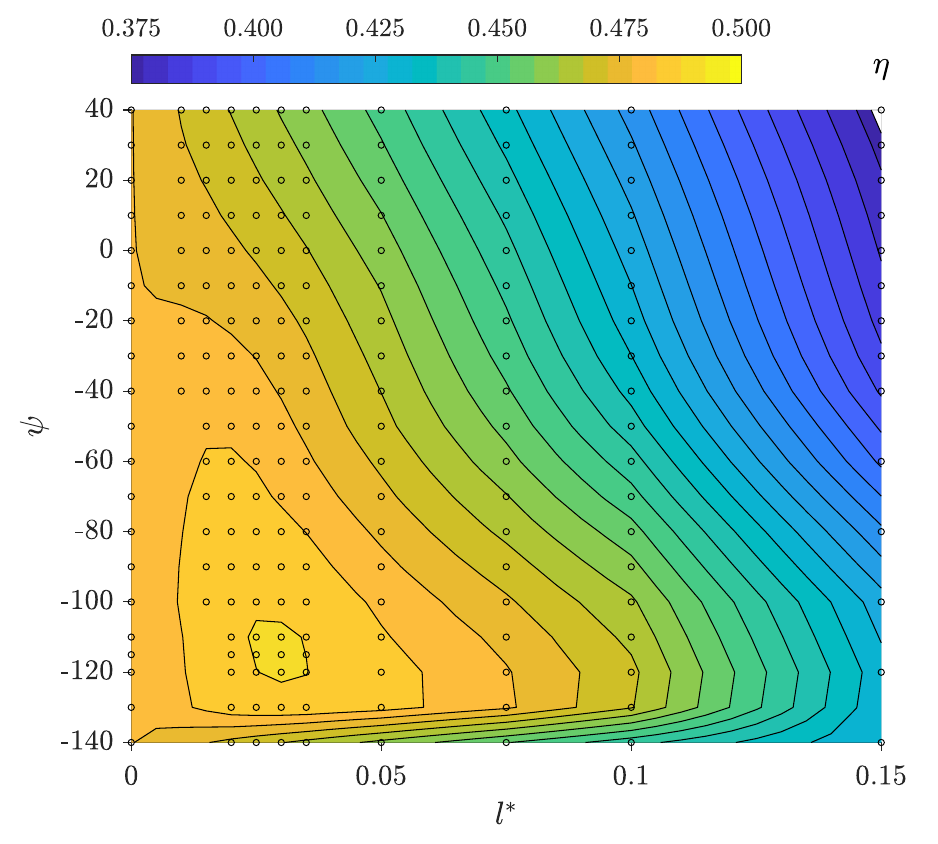}
\caption{Single foil.}
\label{eta_single_plate}
\end{subfigure}
\hfill
\begin{subfigure}[t]{.49\textwidth}
\centering
\includegraphics[width=\linewidth]{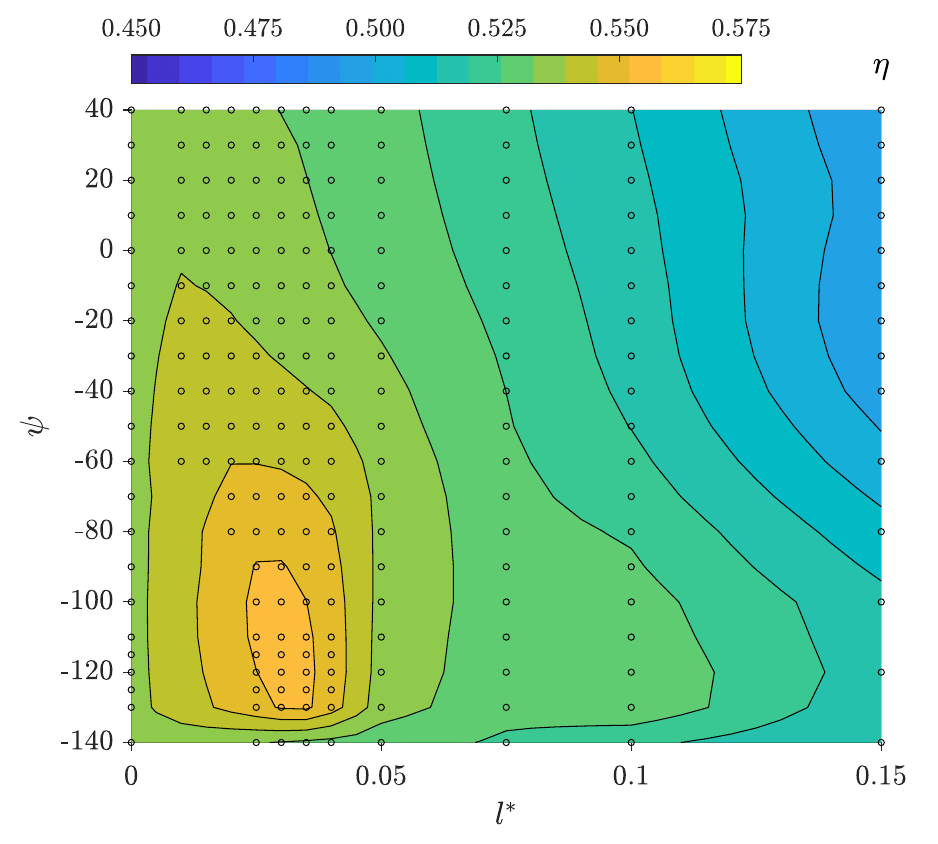}
\caption{Dual foils ($d^*=0.005$).}
\label{eta_dual_plate}
\end{subfigure}

\caption{Efficiency $\eta$ as a function of the deviation amplitude $l^*$ and
phase shift between the deviation and heaving motions $\psi$.}
\label{Colline_eta}
\end{figure*}

\begin{figure*}[!ht]
\centering
\begin{subfigure}[t]{.49\textwidth}
\centering
\includegraphics[width=\linewidth]{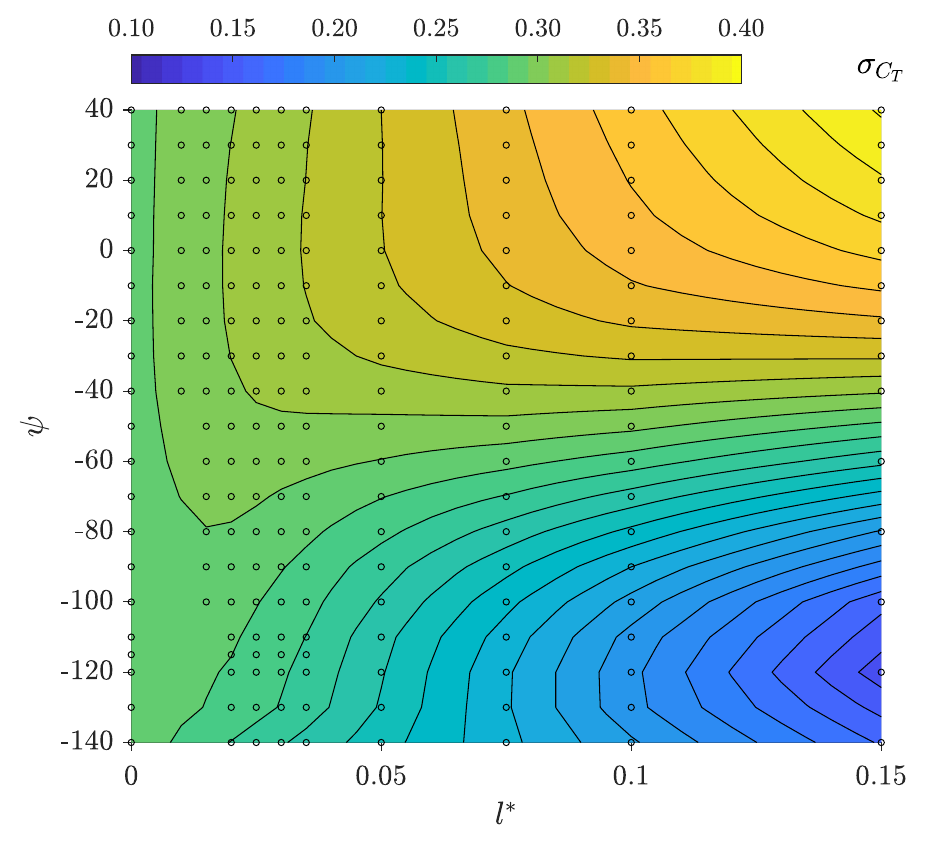}
\caption{Single foil.}
\label{std_single_plate}
\end{subfigure}
\hfill
\begin{subfigure}[t]{.49\textwidth}
\centering
\includegraphics[width=\linewidth]{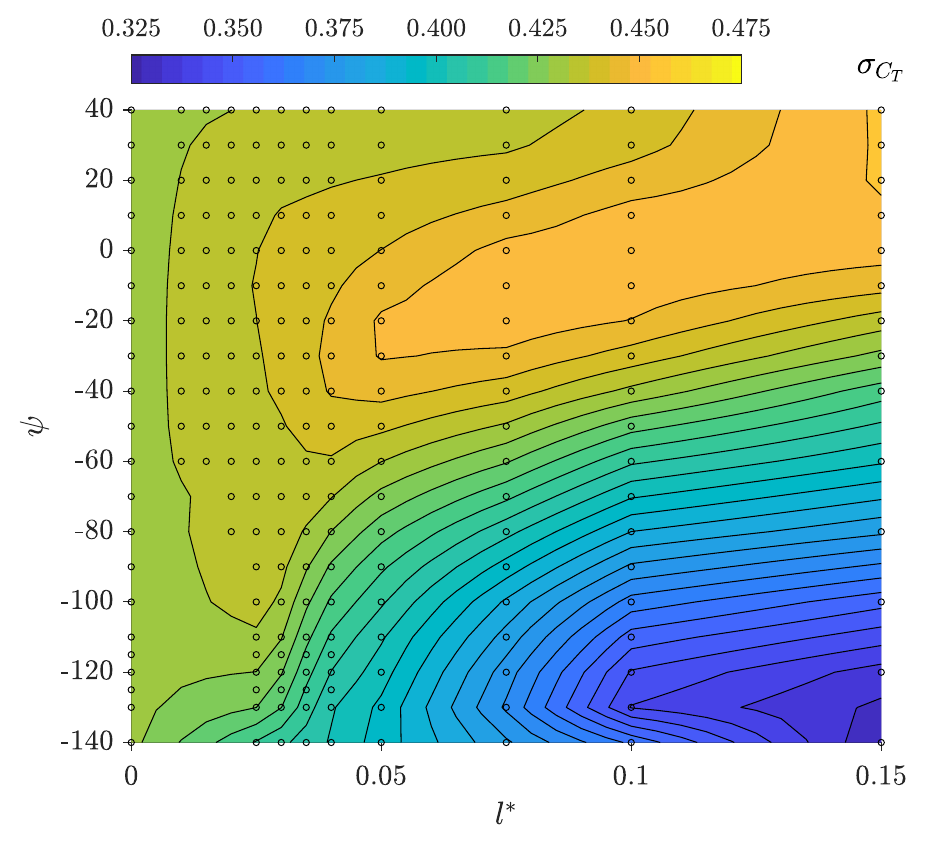}
\caption{Dual foils ($d^*=0.005$).}
\label{std_dual_plate}
\end{subfigure}

\caption{Propulsive standard deviation $\sigma_{C_T}$ as a function of the
deviation amplitude $l^*$ and phase shift between the deviation and heaving
motions~$\psi$.}
\label{Colline_std}
\end{figure*}

The computed performances for cases without deviation based on the metrics
defined in \sect{Sec:perf_metrics} are presented as a function of the phase
shift for some minimum spacings between the foils, see \fig{CT_phase} and
\fig{eta_phase}. Similarly, computed performances for cases with deviation are
presented as a function of the deviation amplitude and phase shift for both
single- and dual-foils in hill charts, see \fig{Colline_CT} to
\fig{Colline_std}.

\fig{CT_phase} shows that the highest thrust occurs at a phase shift of
$\phi=105\degree$ and a minimum spacing of $d^*=0.005$. Lower and higher phase
shifts result in decreased performance, while an increase in the minimum spacing
between the foils results in an overall decrease in thrust for most phase
shifts. At the optimal point, the averaged thrust coefficient is
$\overline{C_T}=0.549$. It is interesting to observe that the maximum thrust in
the case of a single foil (represented here by the minimum spacing
$d^*\rightarrow\infty$) occurs at a different phase shift of $80\degree$ with a
lower thrust coefficient of $\overline{C_T}=0.36$. These results confirm a
significant thrust improvement of up to 52.72\% when using a small spacing
between two flapping foils. \fig{eta_phase} shows the same trend for the
efficiency: the maximum $\eta=0.542$ occurs at the same phase shift
$\phi=105\degree$ and minimum spacing $d^*=0.005$ for the case with two foils.
Interestingly, the optimal efficiency $\eta=0.478$ of the single-foil case does
not occur at the same phase shift as its maximum thrust, but rather at the same
one as the most efficient two-foil case. Lastly, a relative efficiency increase
of up to 13.50\% is observed when two foils are used instead of one.

\begin{figure}[!hb]
\centering
\includegraphics[width=0.98\linewidth]{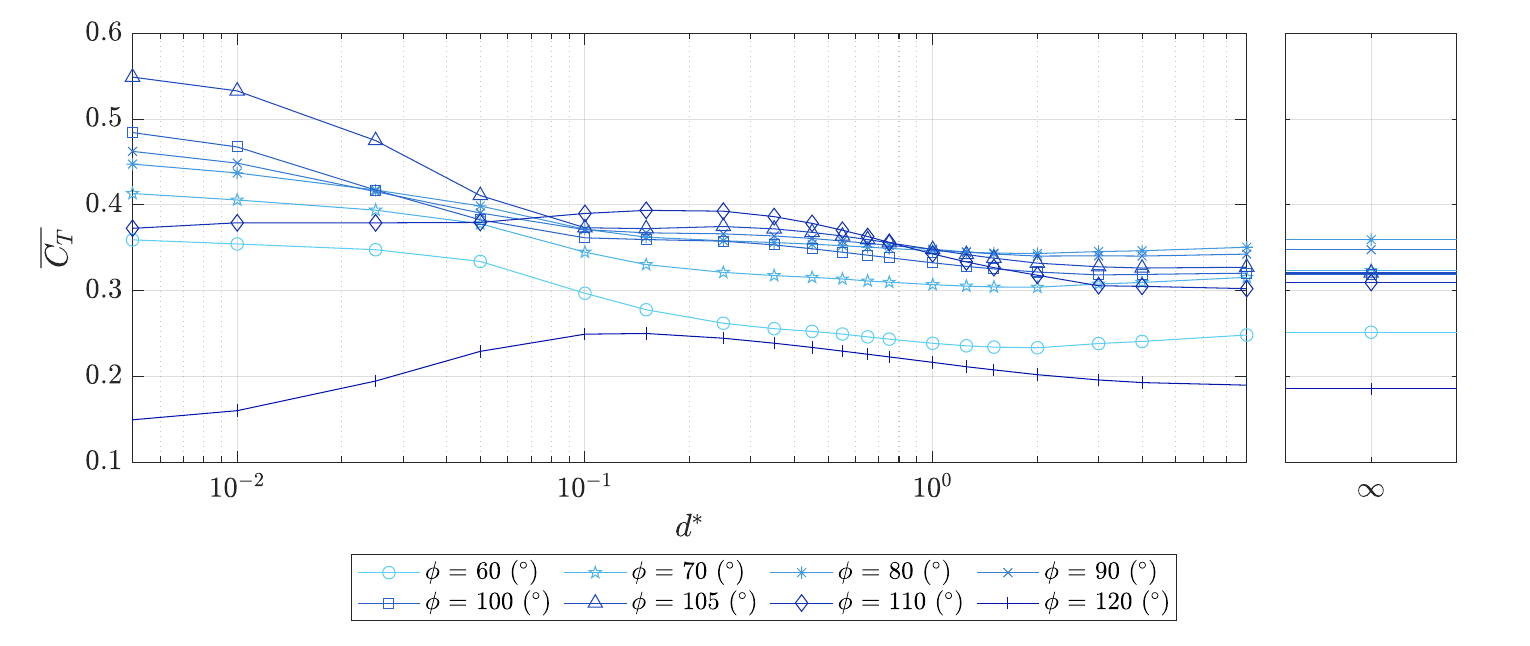} 
\caption{Averaged thrust coefficient $\overline{C_T}$ as a function of the
minimum spacing $d^*$ for different phase shifts~$\phi$ (no deviation).}
\label{CT_gap}
\end{figure}

\begin{figure}[!ht]
\centering
\includegraphics[width=0.98\linewidth]{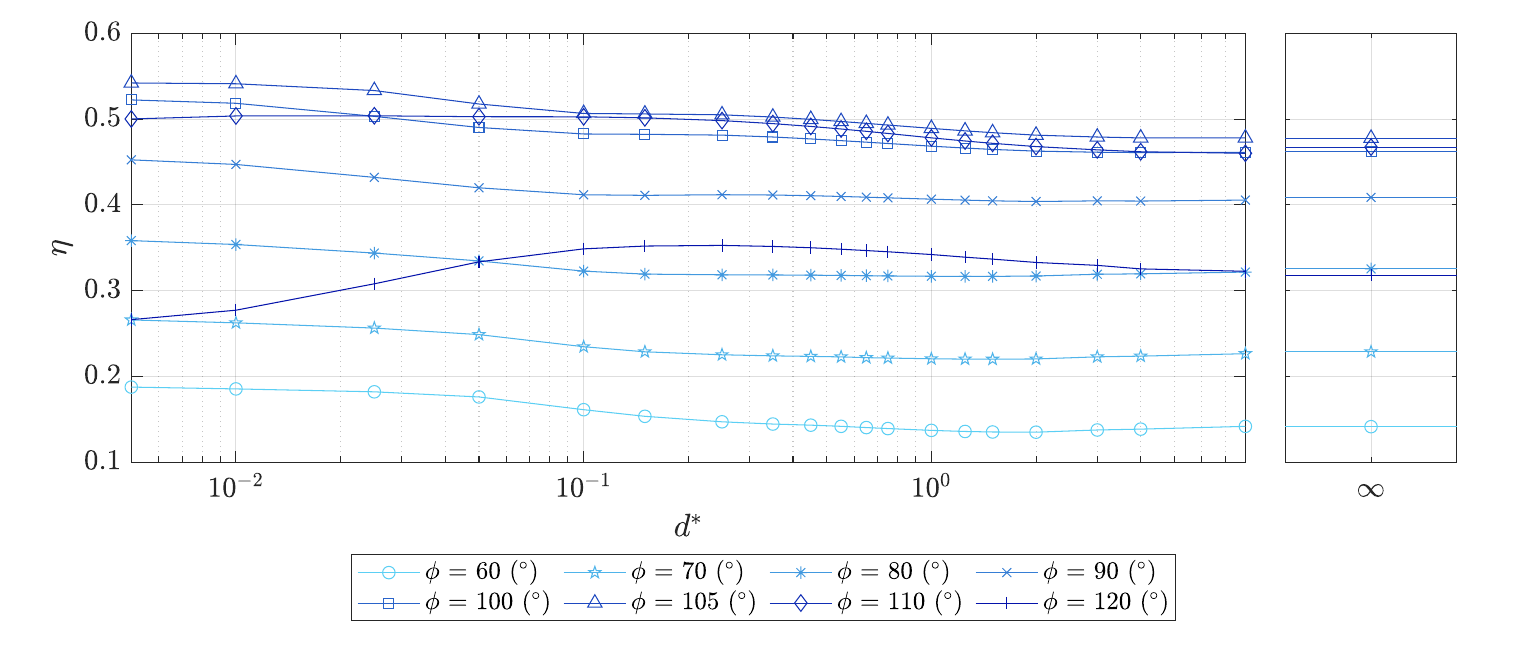} 
\caption{Efficiency $\eta$ as a function of the minimum spacing $d^*$ for
different phase shifts~$\phi$ (no deviation).}
\label{eta_gap}
\end{figure}

To better observe the impact of the minimum spacing between the foils,
\fig{CT_gap} and \fig{eta_gap} illustrate the cycle-averaged performance
metrics as a function of $d^*$ for all the different phase shifts.
As seen from \fig{CT_gap}, the best thrust is found at the smallest minimum
spacing tested ($d^* = 0.005$) and at a phase shift $\phi=105\degree$. However,
as the minimum spacing increases, the curve with $\phi=110\degree$ presents a
higher thrust than the case with $\phi=105\degree$. In \fig{eta_gap}, it can be
seen that the curve associated with the phase shift 105$\degree$ always
presents the best efficiency for all of the minimum spacings $d^*$. At $d^*=1$,
the efficiency increases sharply as the minimum spacing decreases for the phase
shift of 105$\degree$ towards the lowest reported value of $d^*=0.005$. This
demonstrates that a near-contact minimum spacing between the foils is necessary
to benefit from clap-and-fling effects. At minimum spacings of $d^*=8$ and
above, the performances asymptotically reach those of a single foil. Concerning
the best case at $d^*=0.005$ and $\phi=105\degree$, flow still occurs between
the foils as they reach the minimum spacing. Therefore, even if the gap is
relatively small, viscous effects in the boundary layers are not strong enough
to completely block the flow between the foils. As such, the trend observed in
\fig{CT_gap} and \fig{eta_gap} suggests that slight performance improvements
are still possible if the minimum spacing is reduced further. However, cases
with a minimum spacing smaller than $d^*=0.005$ were not computed in this
study. Indeed, a finer mesh with prohibitive computational cost would have been
necessary to accurately capture the flow within the small gap when the foils
reach these minimum spacings, especially near the trailing edges.

\subsection*{Cases with deviation}

Subsequently, the deviation is applied on the case with the optimal motion
parameters identified above. \fig{Colline_CT} shows that for a single foil, the
highest averaged thrust coefficient of $\overline{C_T}=0.351$ occurs at a phase
shift of $\psi=10\degree$ and deviation amplitude $l^* = 0.100$. As shown on the
hill chart, higher deviation amplitudes (near the maximum value considered of
$l^* = 0.150$) tend to increase the thrust when a positive phase shift $\psi$
close to $0\degree$ is used. Interestingly, this is not the case for the dual
foils subjected to the clap-and-fling mechanism where a lower deviation
amplitude and a significant negative phase shift yield better performances.
Indeed, the maximum thrust coefficient $\overline{C_T}=0.610$ occurs at
$\psi=-90\degree$ and $l^* = 0.025$. This value represents a relative increase
of 73.79\% with respect to the highest reported value in single-foil cases,
which is a significant improvement. As for the averaged power coefficient hill
charts presented in \fig{Colline_CP}, similar trends to those for the thrust can
be observed.  Important differences are instead seen on the efficiency hill
chart presented in \fig{Colline_eta} where optimal efficiencies are obtained in
similar regions for both single- and dual-foil cases with lower deviation
amplitudes and large negative phase shifts. For the single-foil configuration,
the maximum efficiency of $\eta=0.488$ occurs at $\psi=-100\degree$ and $l^*
= 0.030$ while for the dual-foil one, it occurs at $\psi=-125\degree$ and $l^*
= 0.035$ with a maximum value of $\eta=0.560$.  Thus, employing two foils
yields a relative efficiency increase of up to 14.72\%. 

From the above discussion, we observe that using two foils subjected to clap
and fling moves the optimal averaged thrust coefficient to a combination of
deviation amplitude and phase shift closer to the one for the best efficiency,
which is desirable. Indeed, the averaged thrust coefficient obtained at
$\psi=-125\degree$ and $l^*$ = 0.035 is $\overline{C_T}=0.563$, which represents
a relative increase of 2.49\% compared to the case with no deviation
($\overline{C_T}=0.549$).  As for the efficiency, the value of 0.560 obtained
with the same parameters represents an increase of 3.21\% compared to the case
without deviation ($\eta=0.542$). Therefore, properly tuning the deviation
motion produces modest increases in both the thrust and efficiency in dual-foil
configurations.

Moreover, it is observed in \fig{Colline_std} that increasing the deviation
amplitude with negative phase shifts $\psi$ below around $-40\degree$ does level
the aerodynamic forces as reported by \cite{bos_influence_2008} for both single-
and dual-foil configurations. It is also observed that the standard deviation
increases with the deviation amplitude for phase shifts higher than around
$-40\degree$. The maximum value of the propulsive standard deviation reaches
$\sigma_{C_T}=0.453$ for the case $l^* = 0.150$ and $\psi=40\degree$,
corresponding to a configuration with the highest deviation amplitude tested in
the study.  Higher values could likely be reached by increasing $l^*$ or $\psi$
further, but this would significantly reduce the thrust and efficiency as
observed in \fig{Colline_CT} and \fig{Colline_eta}, which is undesirable and
thus not tested.  For the optimal efficiency configuration of the dual-foil
case, the reported value of the propulsive standard deviation is
$\sigma_{C_T}=0.412$. This represents a slight decrease compared to the base
case without deviation where $\sigma_{C_T}=0.429$. Therefore, in addition to
increasing efficiency, a negative phase shift of $\psi=-125\degree$ at a
deviation amplitude of $l^*$ = 0.035 also slightly levels the unsteady forces
resulting from the whole flapping motion.  On the other hand, the dual-foil
configuration producing the best thrust ($C_T = 0.610$ with $\psi = -90\degree$
and $l^* = 0.025$) presents a slightly higher propulsive standard deviation of
$\sigma_{C_T}=0.436$, corresponding to more force variations throughout a motion
cycle compared to the case without deviation.  \sect{Sec:Flow_field} and
\sect{Sec:angle_attack} further discuss these aspects by analyzing the flow
field topology against the kinematics of the flapping foils using the deviation
motion.

\subsection{Flow fields topology}
\label{Sec:Flow_field}

\subsection*{Cases without deviation}
\begin{figure}[!ht]
\centering
\includegraphics[width=\linewidth]{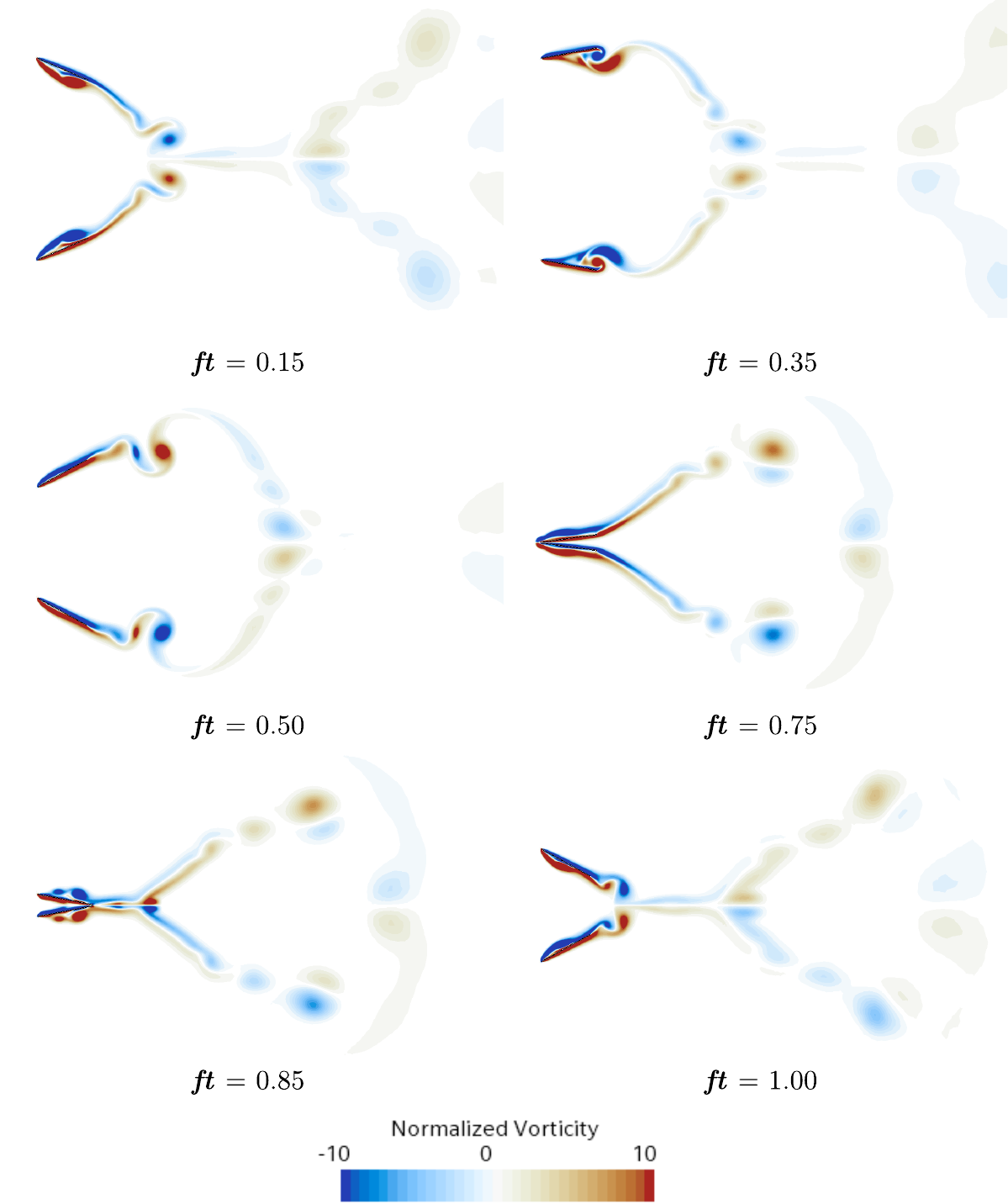} 
\caption{Normalized vorticity field ($\omega c/U_{\infty}$) for
$\phi=105\degree$ and $d^*=0.005$.}
\label{vorticity_double_105_0.005}
\end{figure}

\fig{vorticity_double_105_0.005} shows the vorticity field of the best case
without deviation for thrust and efficiency ($d^*=0.005$ and
$\phi=105\degree$) at several time frames during one motion cycle. As seen in
the figure, flow separation occurs during the clap motion on the boundary
layers of each foil along with the formation of leading-edge vortices.
However, because of the low reduced frequency ($f^{*}=0.15$), each vortex shed
from the leading edge is convected quickly in the wake, which minimizes its
interaction with the foils. This wake behavior is thus in agreement with that
of a single foil as shown by \cite{olivier_fluid-structure_2014}.

In addition, the best thrust and efficiency configurations exhibit moderate
leading-edge vortices. Such vortices in low-Reynolds-number flow regimes are
known to modulate significantly aerodynamic forces because of the low-pressure
zone within their core. In addition, even though the flow separates near the
leading edges to form vortices, it may remain attached at the trailing edges,
maintaining a smooth flow that satisfies the Kutta condition
\citep{sane_aerodynamics_2003}. This phenomenon is known as delayed stall, and
it strongly depends on the motion parameters in flapping foils
\citep{chen_leading-edge_2020}. In the current study, the optimal phase shift of
$\phi=105\degree$ combined with the reduced frequency $f^*=0.15$ and baseline
pitching amplitude $\theta_0=30\degree$ allows the foils to change their course
before actual stall occurs. In the meantime, the foils benefit from high thrust
because of their high angle of attack.

\begin{figure}[tb]
\centering
\includegraphics[width=\linewidth]{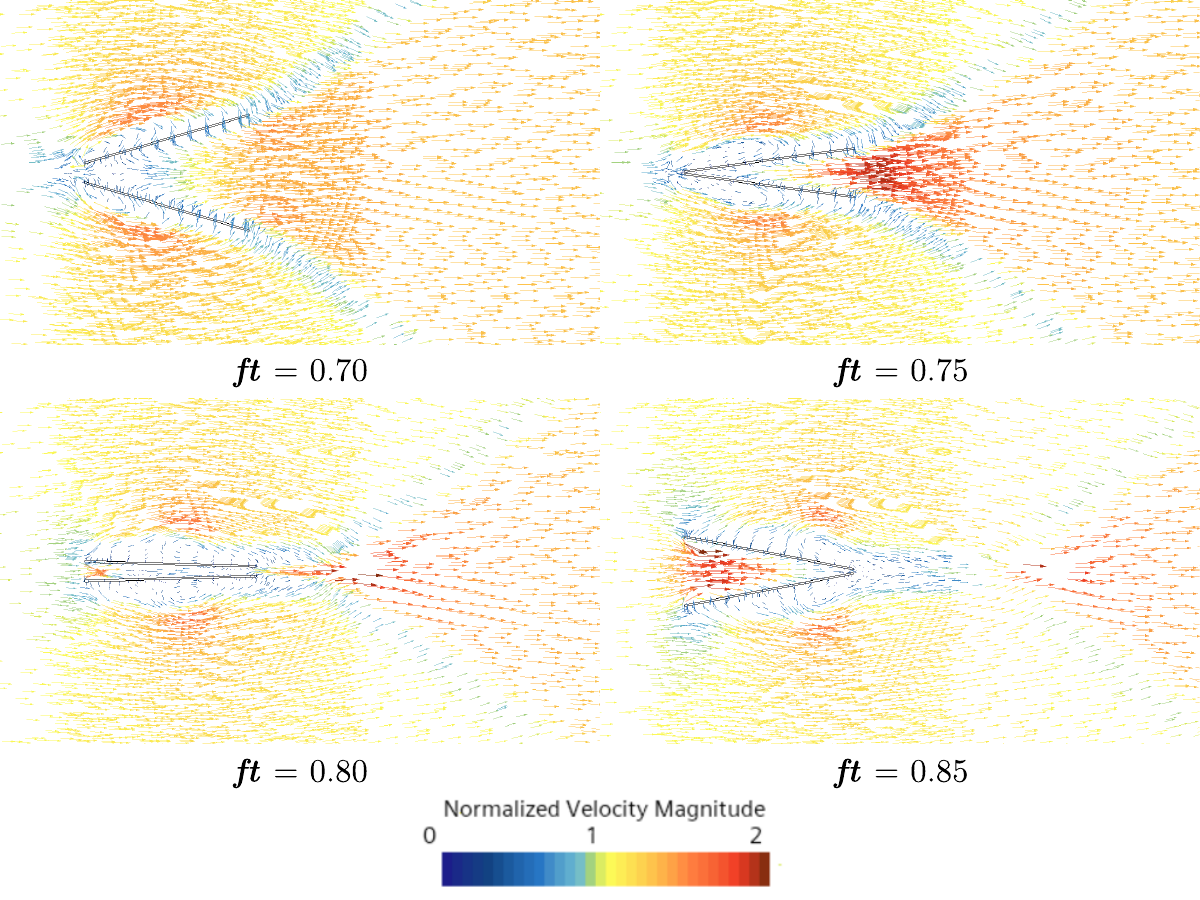} 
\caption{Velocity vector field for $\phi=105\degree$ and $d^*=0.005$.
The vector length is scaled with the normalized velocity magnitude.}
\label{velocity_double_105_0.005}
\end{figure}

Furthermore, it is observed in \fig{vorticity_double_105_0.005} that the minimum
spacing occurs at the trailing edge during the motion. This is a key aspect of
the clap-and-fling mechanism since the flow between the two foils is mainly
responsible for generating thrust improvements. Indeed, when the minimum spacing
is reached at the trailing edge at $ft=0.85$, the fluid between the foils is
ejected downstream at high speed. This build-up of high velocity can be better
observed in \fig{velocity_double_105_0.005}, which presents the instantaneous
velocity vector field. As stated by \cite{shyy_aerodynamics_2007}, this fast
flow ejection allows the trailing edge vortex of each wing to be shed more
quickly in the wake, which mitigates the Wagner effect where a latency typically
occurs in the establishment of circulation of the foils' bound vortices
counteracted by the trailing edge vortices. As a result, the thrust generation
resulting from this quick circulation buildup is greatly accelerated. As for the
subsequent fling phase, fluid is inhaled upstream of the foils into the gap
created as they separate, greatly increasing circulation. This causes strong
vortex generation at their leading edges on the inner surfaces, as seen at
$ft=1.00$ in \fig{vorticity_double_105_0.005}. This fast circulation of opposite
signs buildup on each foil generates high thrust on the two foils. This
mechanism is analogous to a ground effect observed in various aerodynamics
applications.

\begin{figure}[!ht]
\centering
\includegraphics[width=\linewidth]{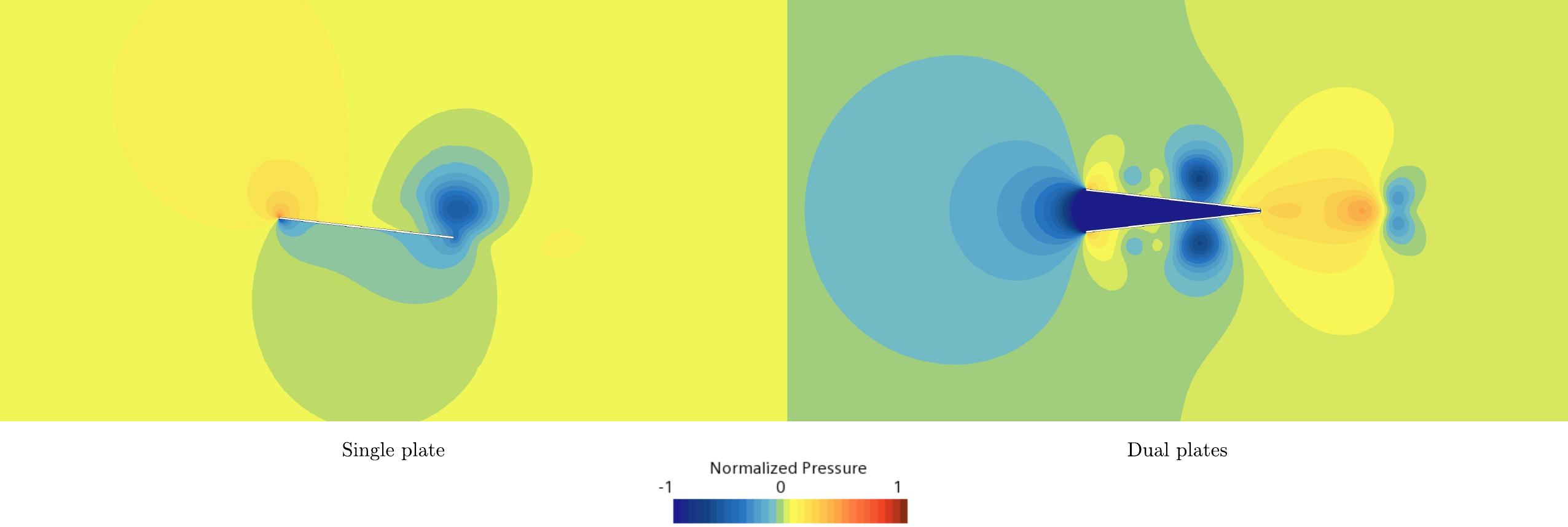} 
\caption{Comparison of the normalized pressure field ($p/(\rho U_{\infty}^2)$)
of the single- and dual-foils cases with $\phi=105\degree$ and $d^*=0.005$ at $ft=0.826$. }
\label{pressure_fields_ground_effect}
\end{figure}

\begin{figure}[!ht]
\centering
\includegraphics[width=\linewidth]{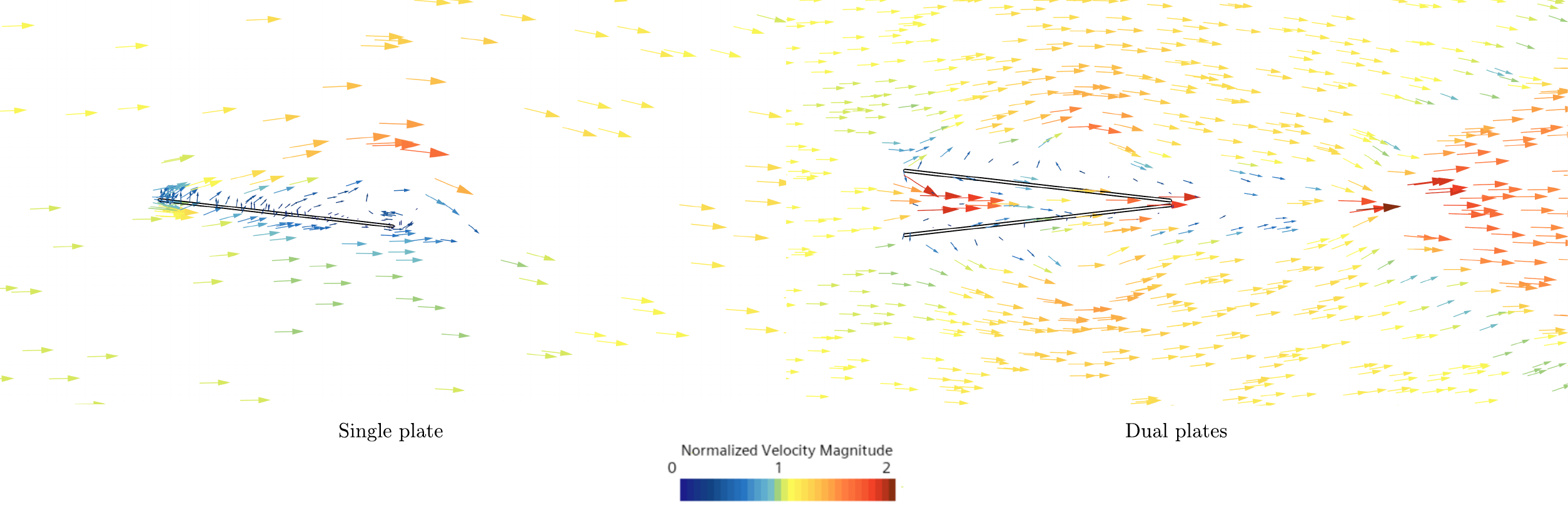} 
\caption{Comparison of the velocity vector field for the single- and
dual-foil configurations with $\phi=105\degree$ and $d^*=0.005$ at $ft=0.826$.
The vector length is scaled with the normalized velocity magnitude. }
\label{vector_fields_ground_effect}
\end{figure}

In the case of flapping wings undergoing clap and fling, a ground-effect-like
phenomenon occurs when the foils reach proximity during the clap phase.  This is
shown in \fig{pressure_fields_ground_effect} and
\fig{vector_fields_ground_effect} where a comparison of the normalized pressure
field and velocity vector field for the best dual-foil case with
$\phi=105\degree$ and $d^*=0.005$ with the equivalent single-foil case is
presented at $ft=0.826$ (when the minimum spacing is reached at the trailing
edge). The presence of the second foil produces a symmetry where streamlines are
deviated so that they are parallel to the symmetry plane. Indeed, this symmetry
is similar to that of a solid boundary if we neglect viscous effects and thus
confine the flow between the two foils, changing its behavior. This is observed
in the velocity vector field where the fluid begins to accelerate between the
foils as they start to separate during the fling phase of the motion. Through
the latter and with the inclination of the foils, a ground effect occurs and
greatly diminishes the already present low-pressure zone observed on the lower
surface of the single foil. Therefore, the force distributions are affected,
which predominantly increases the thrust.

\begin{figure}[!ht]
\centering
\includegraphics[width=\linewidth]{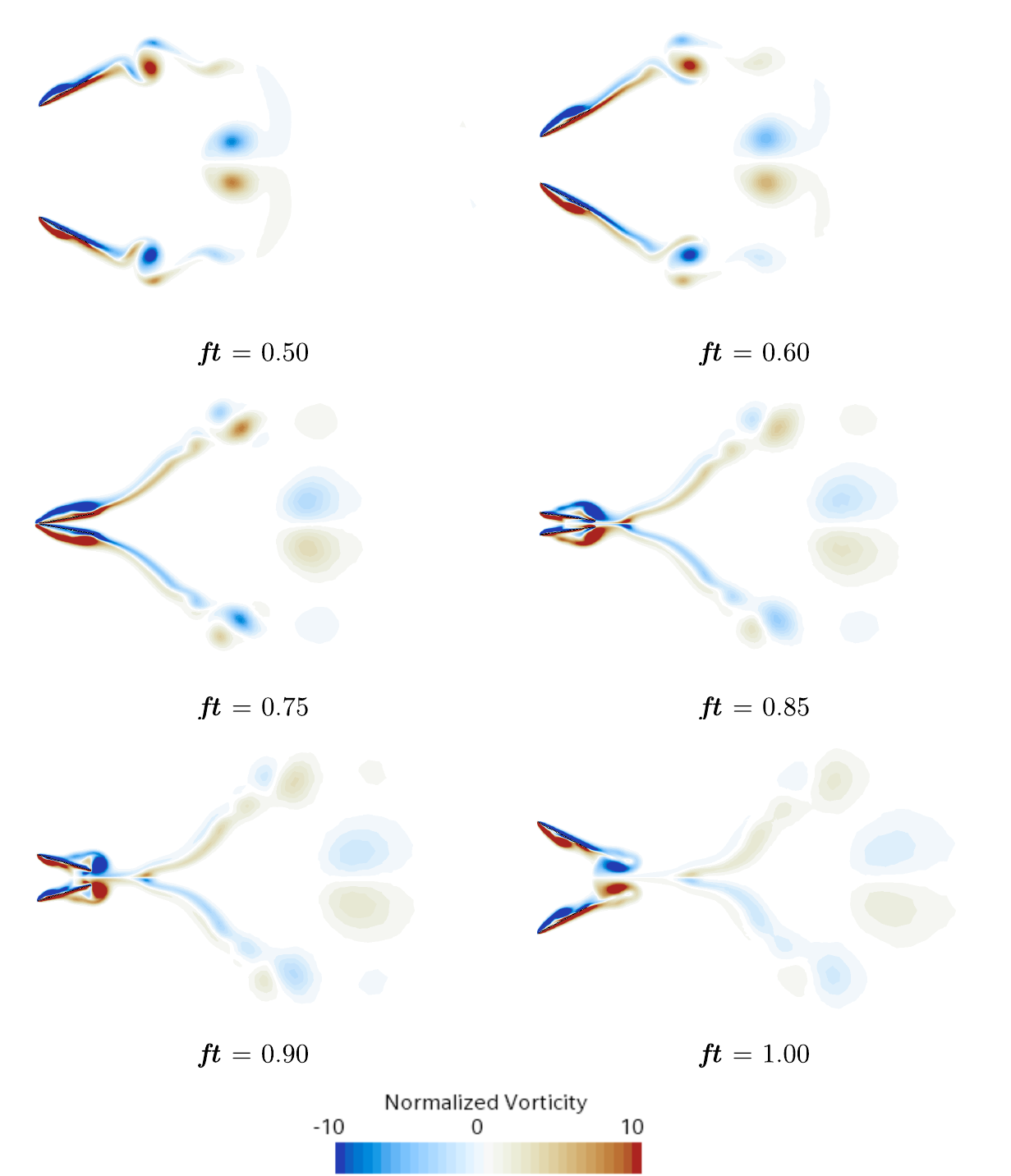} 
\caption{Normalized vorticity field ($\omega c/U_{\infty}$) for
$\phi=110\degree$ and $d^*=0.005$.}
\label{vorticity_double_110_0.005}
\end{figure}

\begin{figure}[!ht]
\centering
\includegraphics[width=\linewidth]{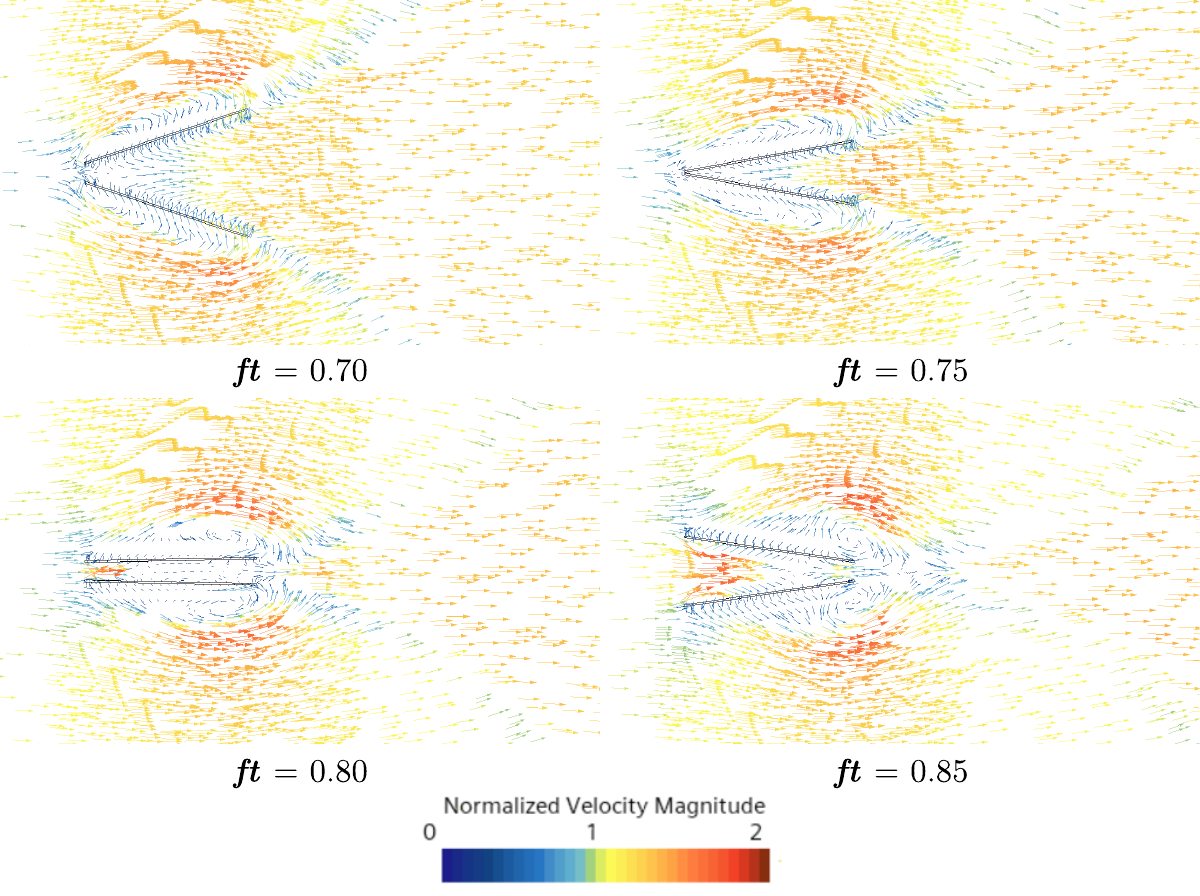} 
\caption{Velocity vector field for $\phi=110\degree$ and $d^*=0.005$.
The vector length is scaled with the normalized velocity magnitude.}
\label{velocity_double_110_0.005}
\end{figure}

To better understand the impact of the kinematics on the clap-and-fling
performance improvement, vorticity and velocity fields of another case are
presented in \fig{vorticity_double_110_0.005} and
\fig{velocity_double_110_0.005}. It can be seen that the change in phase shift
significantly impacts the kinematics of the foils as the minimum spacing occurs
at the leading edge at $ft=0.75$ for this case. This configuration blocks the
fluid upstream of the foils from being injected in between and thus reduces its
thrust and efficiency. Since the minimum spacing now occurs at the leading edge,
the boundary layers detach around $ft=0.85$ when the spacing increases. The
fluid is then convected downstream at a lower velocity when compared to the
optimal case. Indeed, \fig{velocity_double_110_0.005} also shows high-velocity
vectors upstream of the foils near their leading edges during the clap phase,
which is detrimental to the thrust. The latter is thus consequently lower, and
the trailing edge vortices of each wing are convected more slowly. These
observations emphasize the importance of the phase shift, particularly at small
minimum spacings. If not properly chosen, the flow can be stuck between the
foils, resulting in suboptimal thrust performances. A phase shift of
$\phi=105\degree$, which is higher than the typical case of $\phi=90\degree$, is
necessary to increase the performances. However, phase shifts larger than
$105\degree$ impede the flow by changing the chordwise position of the minimum
spacing between the foils.

\begin{figure}[!ht]
\centering
\includegraphics[width=\linewidth]{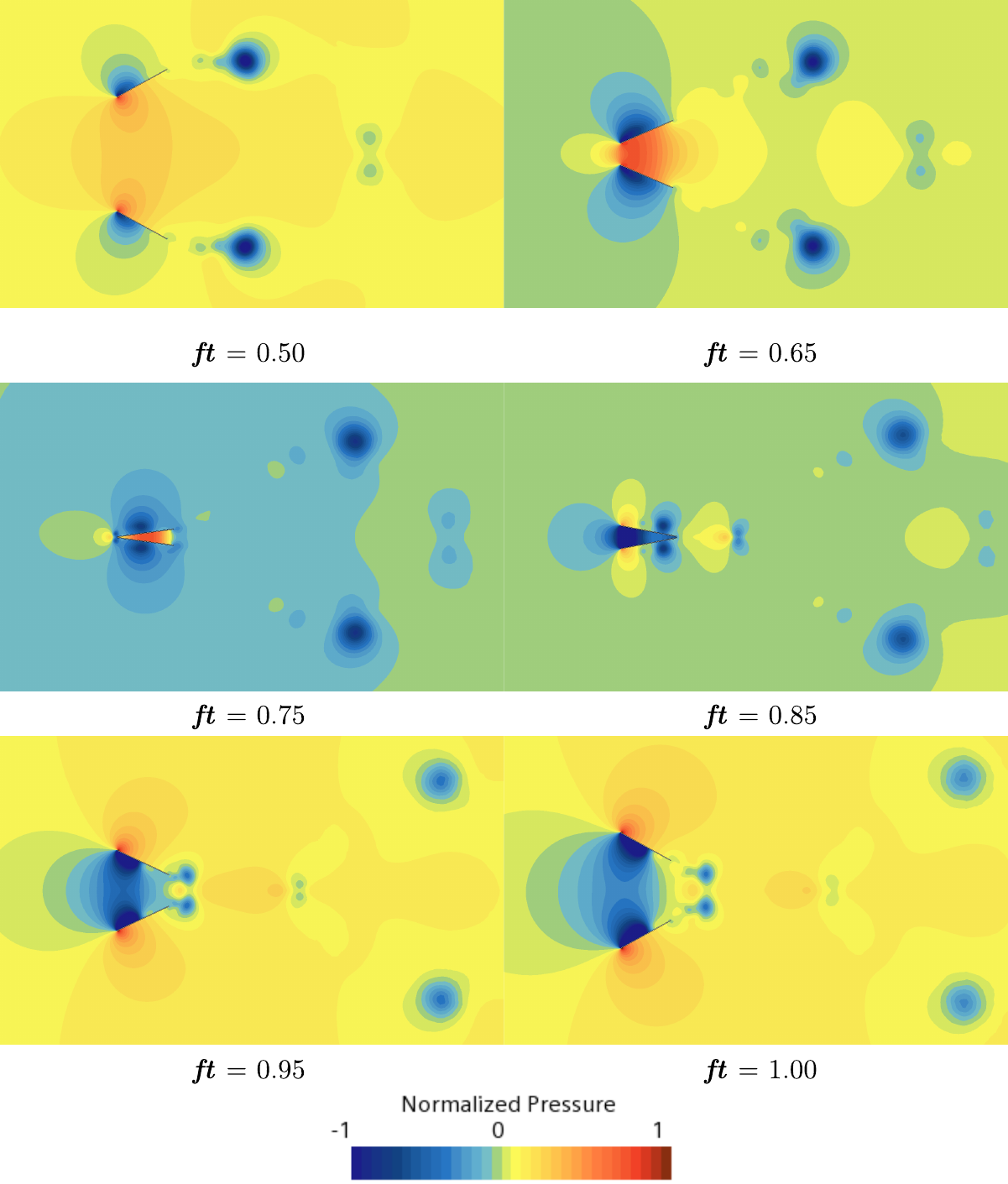} 
\caption{Normalized pressure field ($p/(\rho U_{\infty}^2)$) for
$\phi=105\degree$ and $d^*=0.005$.}
\label{pressure_105}
\end{figure}

\begin{figure}[!tb]
\centering
\includegraphics[width=\linewidth]{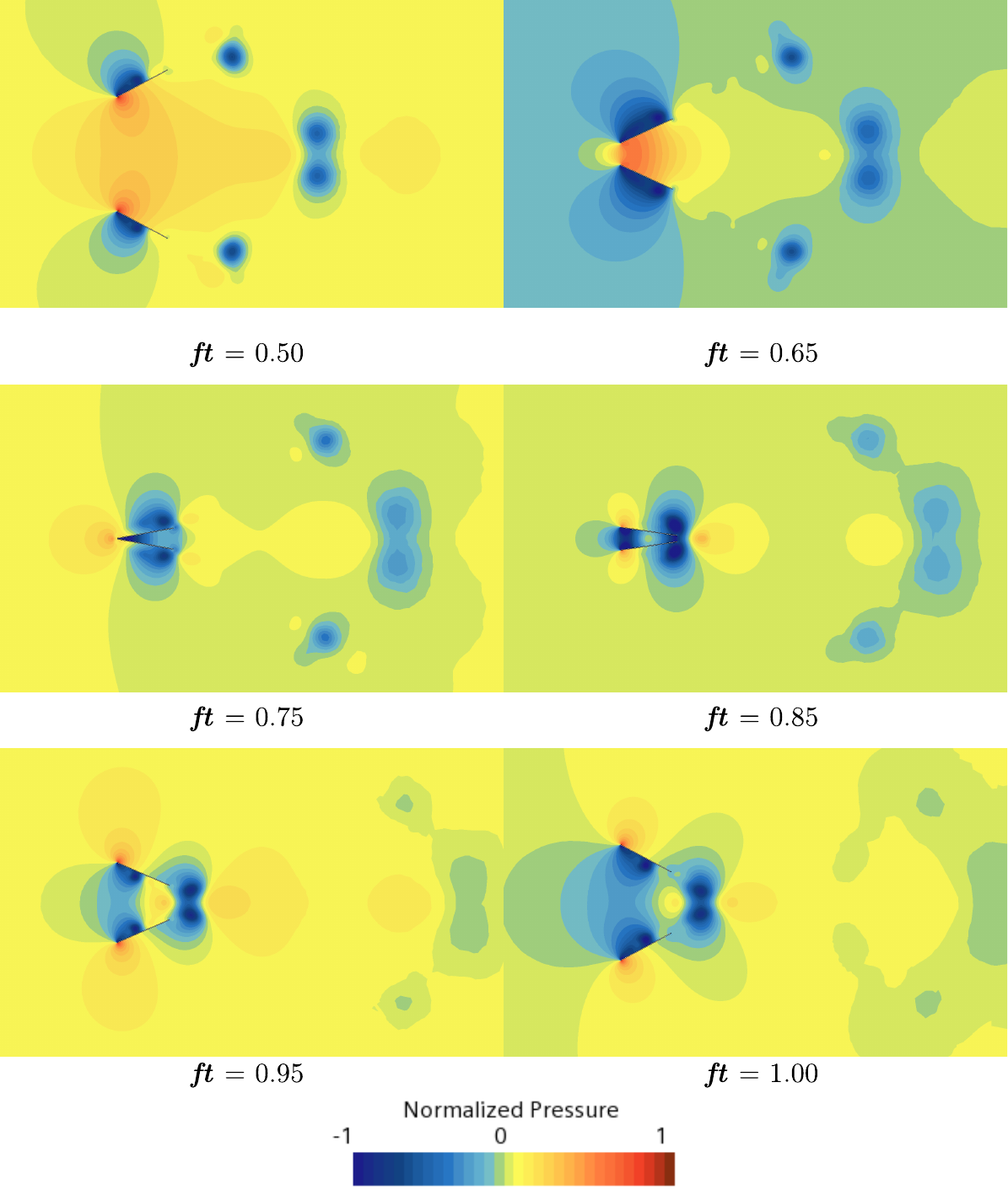} 
\caption{Normalized pressure field ($p/(\rho U_{\infty}^2)$) for
$\phi=110\degree$ and $d^*=0.005$.}
\label{pressure_110}
\end{figure}

To support this discussion further, static pressure contours for the two
previous cases are presented respectively in \fig{pressure_105} and
\fig{pressure_110}. For the optimal case with $\phi=105\degree$, it can be seen
in \fig{pressure_105} that a high positive static pressure zone is present to
the right (downstream) of the foils during the clap phase, which tends to push
the foils to the left and thus contribute to the thrust. As the foils move
closer to each other, this positive pressure only increases with a negative
pressure zone building to the left of the foils as their inclination changes.
As the foils undergo the clap-and-fling motion, this low-pressure zone only gets
larger in size and magnitude thus pushing once again the foils in the direction
opposite to the flow.

For the case with $\phi=110\degree$ presented in \fig{pressure_110}, the
positive pressure zone during the clap phase is not as high in magnitude and
thus results in a lower thrust. Moreover, low pressure develops between the
foils after they reach their minimum spacing at the leading edge. At the same
time, the vortex core on the outer side of each foil produces a low-pressure
zone near the trailing edge (see \fig{vorticity_double_110_0.005} between time
$ft=0.85 - 1.00$). The resulting pressure gradient between these two
low-pressure zones is thus very small, which produces low instantaneous thrust
subsequently. Additionally, the closing gap at the leading edge at $ft=0.75$
produces a stagnation effect that increases the pressure at this location,
producing a local force that counteracts the thrust.

\subsection*{Cases with deviation}
\begin{figure}[htb]
\centering
\includegraphics[width=\linewidth]{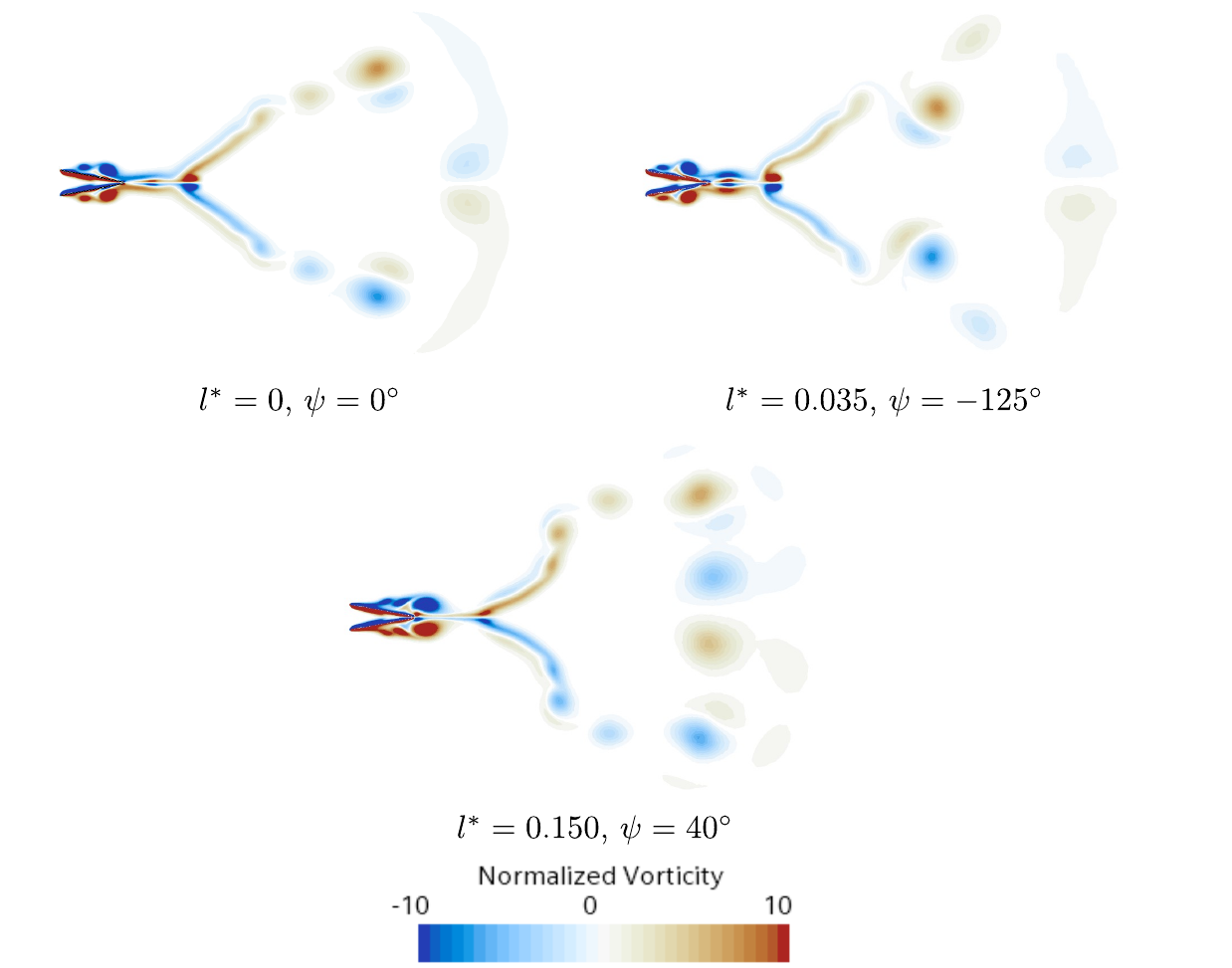} 
\caption{Normalized vorticity field ($\omega c/U_{\infty}$) of the two foils
for different combinations of $\psi$ and $l^*$ at $ft=0.85$.}
\label{vorticity_fields}
\end{figure}

Including the deviation motion has effects on the flow field to some extent.
Firstly, the pressure and vorticity fields of several cases of interest based on
the results discussed in \sect{Sec:parametric_study} are presented. These fields
are shown at $ft=0.85$, which was identified in \sect{Sec:results} as a key
instant where the trailing edges of the foils reach the minimum spacing $d^*$
and generate a significant thrust peak. The basis of comparison is the
configuration with no deviation ($l^* = 0$, $\psi=0\degree$) and is first
compared to the case including deviation with the best efficiency ($l^* =
0.035$, $\psi=-125\degree$). Since this deviation amplitude is moderate, there
are no significant differences between the two cases in the flow fields during
one period of motion, which agrees with similar results by
\cite{jung_role_2024}. The performance improvements will thus be explained by a
different approach in \sect{Sec:angle_attack}. The last case of interest with
$l_0 = 0.150$ and $\psi=40\degree$ represents a configuration with substantial
deviation where the propulsive standard deviation $\sigma_{C_T}=0.453$ is the
highest in the range of $l^*$ considered in this parametric study. For such a
configuration, important differences are noticeable in the flow fields. Indeed,
\fig{vorticity_fields} shows that this case exhibits important flow separation
on both flapping foils and more leading-edge vortices.  These are eventually
shed in the wake as shown in \fig{vorticity_fields} where the strongest visible
vortices are being ejected near the trailing edges of the foils. 

\begin{figure}[htb]
\centering
\includegraphics[width=\linewidth]{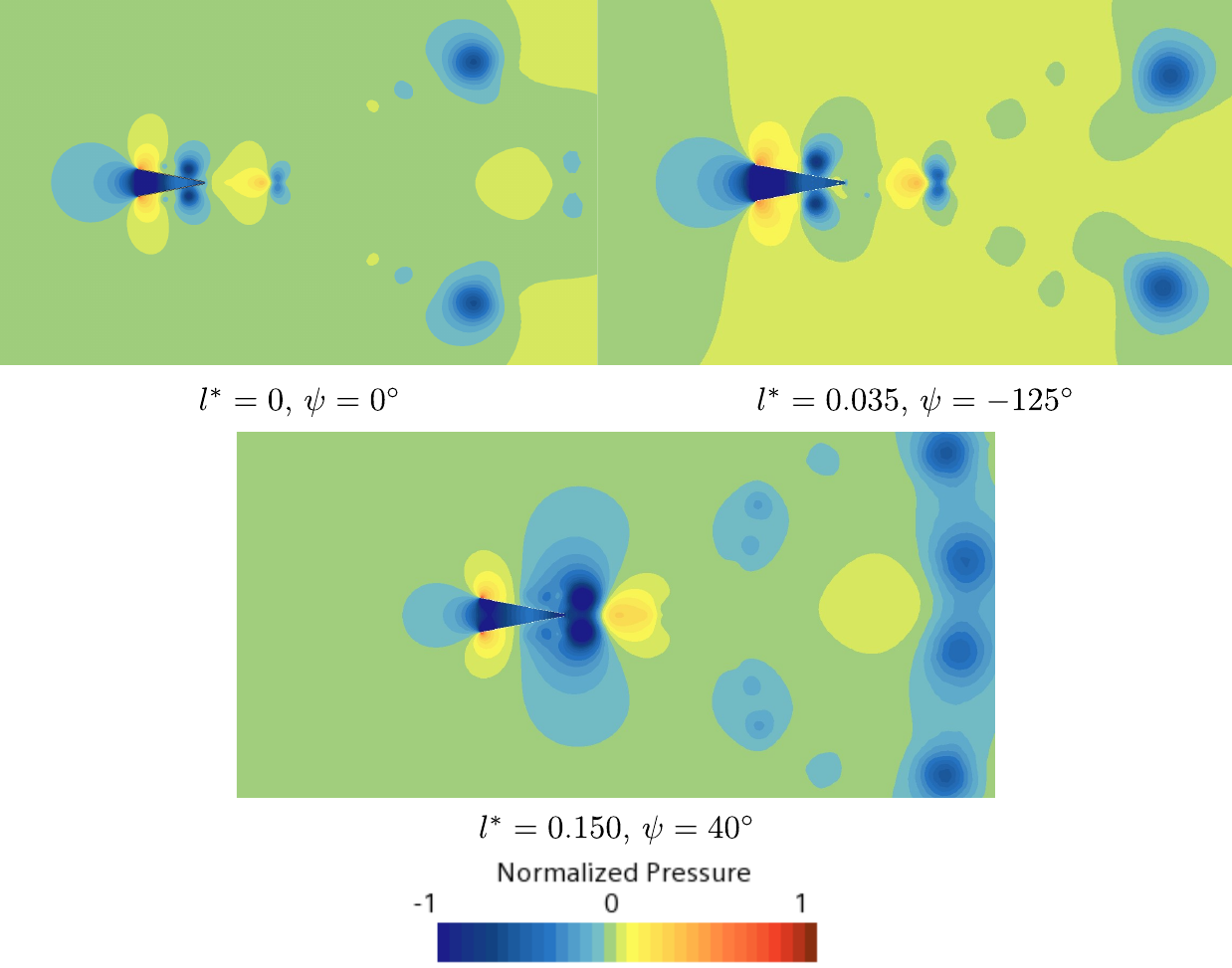} 
\caption{Normalized pressure field ($p/(\rho U_{\infty}^2)$) of the two foils
for different combinations of $\psi$ and $l^*$ at $ft=0.85$.}
\label{pressure_fields}
\end{figure}

Leading-edge vortices are known to modulate significantly aerodynamic
performances due to the low-pressure zone contained within their core
\citep{sane_aerodynamics_2003, chen_dual-stage_2023}. In the cases with no
deviation ($l^* = 0$) and moderate deviation ($l^* = 0.035$,
$\psi=-125\degree$), these leading-edge vortices are moderate and are formed with
proper timing, producing high thrust. In the case with significant deviation
($l^* = 0.150$, $\psi=40\degree$), flow separation and generation of several
leading-edge vortices prevent the flow from remaining attached at the trailing
edges, causing stall to occur. This is most noticeable when the foils are close
to each other at their trailing edges around the instant $ft=0.85$, as shown in
\fig{vorticity_fields}.
The stronger leading-edge vortices generated in this
case create an important low-pressure zone downstream of the foils as observed
in \fig{pressure_fields},
which counteracts the other favorable low-pressure zone upstream of the foils
produced by the clap-and-fling motion. This significantly reduces thrust and
demonstrates that configurations with large deviation may severely impair the
evolution of the flow and decrease the overall performance while increasing
propulsive standard deviation $\sigma_{C_T}$ if the deviation parameters are not
properly chosen.

\subsection{Instantaneous effects on force and power coefficients}
\label{Sec:improvements_two_plates}
So far, the best cases of thrust and efficiency have been established against
the minimum spacing between the foils $d^*$ and the phase shift $\phi$.
However, \fig{CT_gap} and \fig{eta_gap} introduced earlier also present the
thrust coefficient and efficiency of single-foil cases, which correspond to the
points labeled with $d^*\to \infty$. As defined in \sect{Sec:perf_metrics}, the
thrust and power coefficients for the dual-foil simulations are averaged
according to \eq{Eq_Ct_glob} and \eq{Eq_Cp_glob} to obtain an equivalent
performance metric to a single-foil case. As seen in these figures, the curves
for each phase shift reach an asymptote as $d^*$ increases towards the case of
the single foil. Indeed, the thrust coefficient and efficiency are almost
identical at $d^* = 8$ and $d^* \to \infty$, which provides a valuable
indication of the distance required to produce significant clap-and-fling
effects. 

\begin{figure}[!ht]
\centering
\includegraphics[width=\linewidth]{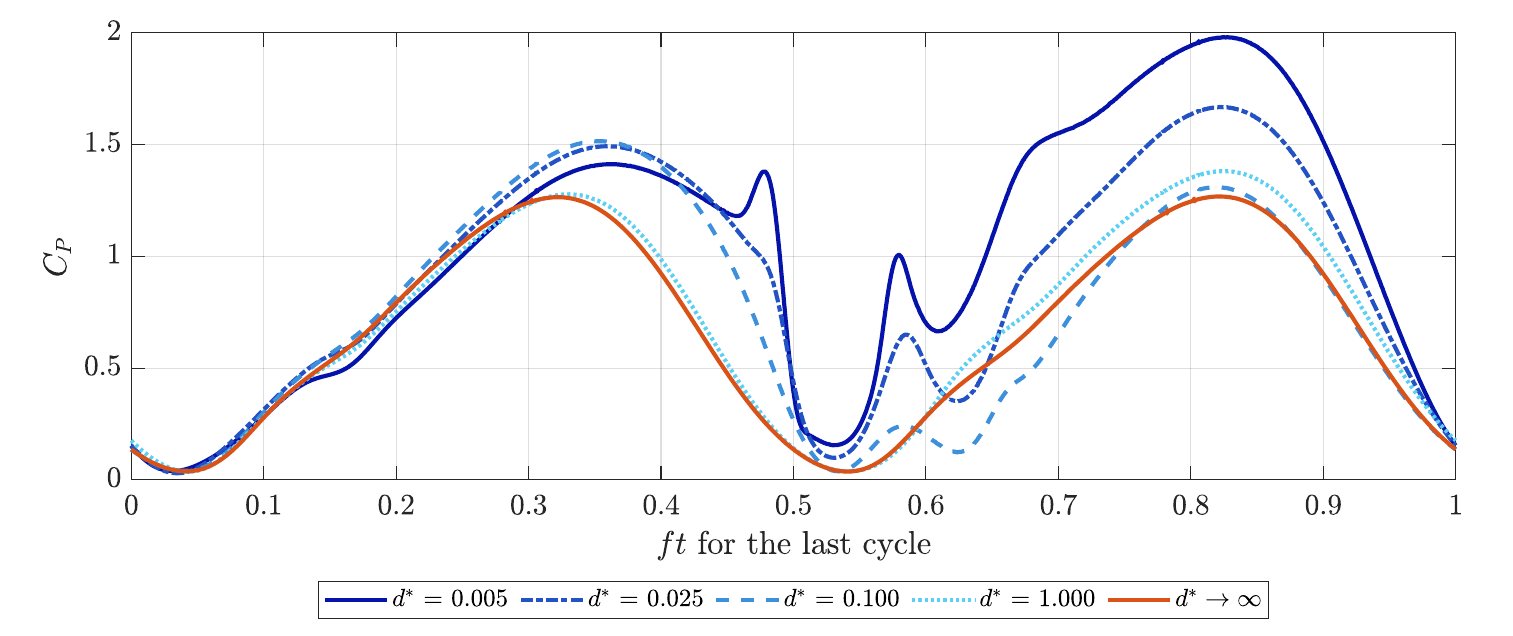}
\caption{Comparison of the power coefficient $C_P$ for $\phi=105\degree$
between the single-foil case and several dual-foil cases.}
\label{CP_comp_100}
\end{figure}

\begin{figure}[!ht]
\centering
\includegraphics[width=\linewidth]{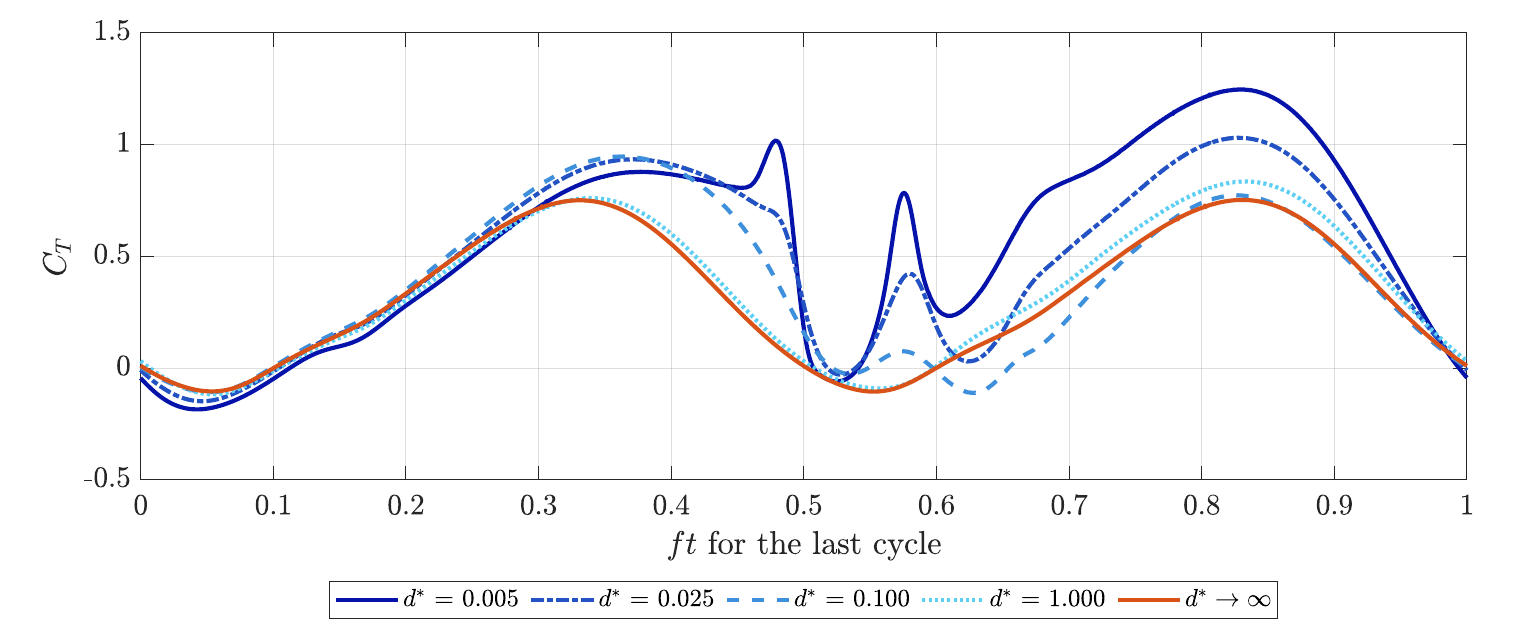}
\caption{Comparison of the thrust coefficient $C_T$ for $\phi=105\degree$
between the single-foil case and several dual-foil cases.}
\label{CT_comp_100}
\end{figure}

Plots of the instantaneous thrust and power coefficient during one cycle are now
compared to assess the performance improvements obtained by using two foils
subjected to the clap-and-fling phenomenon. As seen in \fig{CP_comp_100} and
\fig{CT_comp_100}, using two foils generally increases both the required power
and the thrust. For the dual-foil simulations, the maximum power and thrust
obtained during the downstroke and the upstroke of a given foil are indeed
different. \fig{CT_comp_100} shows this maximum thrust of ${C_T} = 0.75$ at
$ft=0.08$ for the single-foil case and ${C_T} = 1.25$ at $ft=0.078$ for the
best dual-foil case. These thrust peaks are caused by the foils' acceleration
and occur during the upstroke of the foils after the stroke reversal performed
during the clap phase of the previous motion cycle.  Indeed, at these specific
times in the cycle, strong vorticity layers are produced in a short amount of
time. This results in a large rate of change of fluid impulse and, in turn,
increases the thrust \citep{sun_unsteady_2002}.

After the second stroke reversal of the foils observed around $ft=0.30$,
several other thrust peaks are observed in \fig{CT_comp_100}. The most expected
one has a thrust coefficient of ${C_T} = 0.78$ and occurs at $ft=0.826$ as the
foils get close to each other while the fluid is quickly convected downstream
in the wake. This corresponds to when the foils reach their minimum spacing
$d^*$ at the trailing edge. At the same time, the required power also peaks at
${C_P} = 1.01$. Moreover, an interesting observation is a local peak of power
and thrust observed during the clap phase near $ft=0.729$. This corresponds to
the instant where the leading edges of the foils are at their closest distance
to each other, with a large positive pressure zone between them immediately to
the right of the leading edges. The instantaneous performance metrics obtained
at this peak are ${C_P} = 1.38$ and ${C_T} = 1.02$. Interestingly, this instant
represents the second-highest thrust peak encountered during the motion while
the power does not increase as much. Thus, this results in an overall increase
in the instantaneous contribution to efficiency which, in turn, improves the
cycle-averaged performance metrics and highlights the benefits of moving the
foils close to each other during their motion. As mentioned, a proper phase
shift ensures the minimum spacing $d^*$ between the foils is located at their
trailing edges.

\subsection{Deviation kinematics}
\label{Sec:angle_attack}

\begin{figure}[b]
\centering
\includegraphics[width=0.7\linewidth]{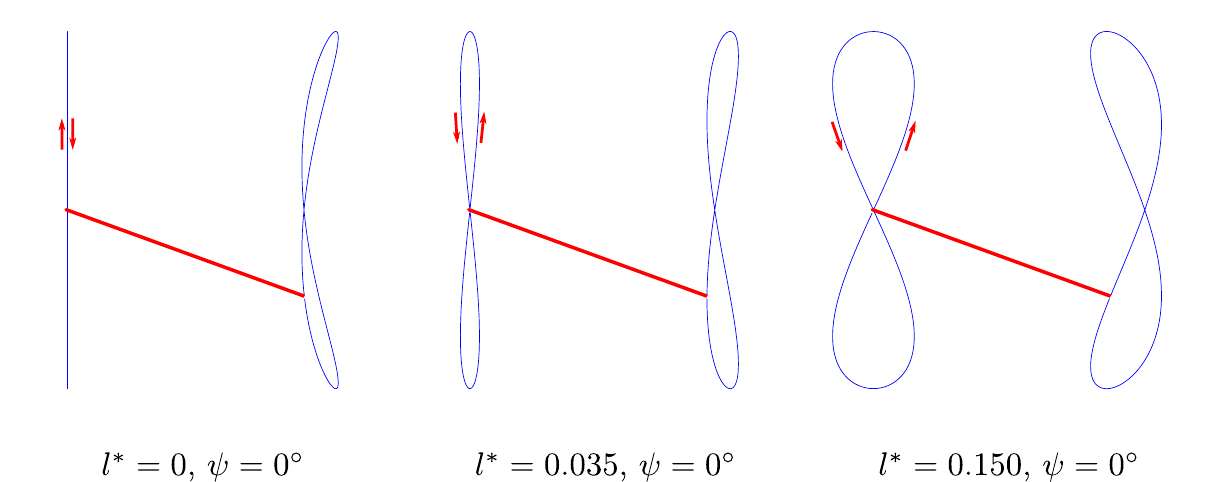} 
\caption{Baseline deviation motion ($\psi=\degree$) for several amplitudes $l^*$.}
\label{deviation_amplitude}
\end{figure}

\begin{figure}[t]
\centering
\includegraphics[width=0.7\linewidth]{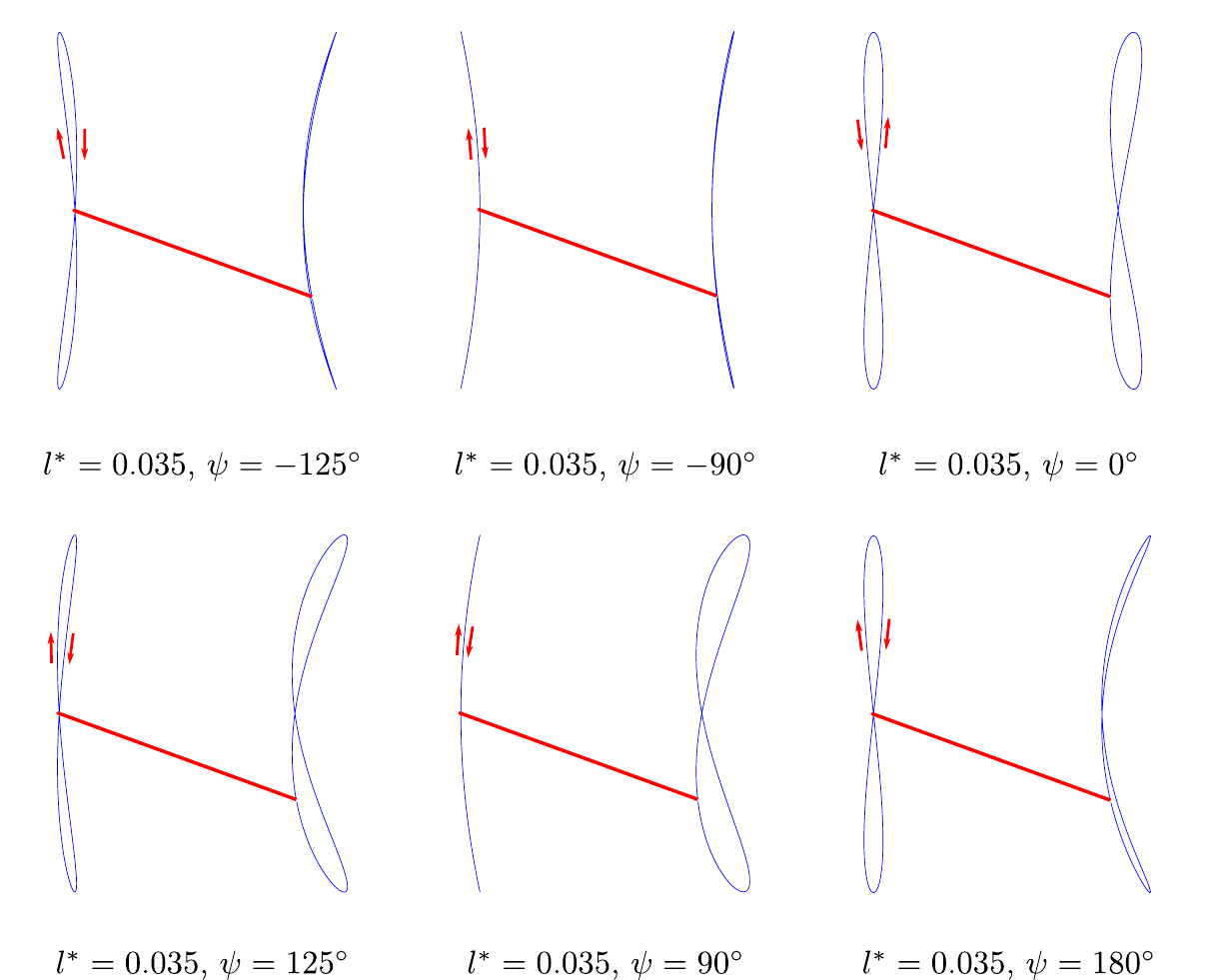} 
\caption{Deviation motion based on the best efficiency configuration
($l^*=0.035$) for several phase shifts $\psi$.}
\label{deviation_phase_shift}
\end{figure}

Based on the results presented in \sect{Sec:parametric_study} and
\sect{Sec:Flow_field}, the kinematics of several cases with deviation are
investigated. More specifically, it will be determined how different
combinations of deviation amplitude $l^*$ and phase shift between the deviation
and heaving motion $\psi$ affect the instantaneous angle of attack of the foils
which necessarily impacts the performances.  \fig{deviation_amplitude} and
\fig{deviation_phase_shift} present the motion of a single foil according to
different combinations of the parameters $l^*$ and $\psi$. The trajectory of the
pivot point position located
near the leading edge at $x_P^*=0.005$ as well as the point located at the
trailing edge are represented by the blue curves for one period of motion. As
seen in \fig{deviation_amplitude}, the pivot point of a flapping foil with no
deviation moves only vertically according to the heaving motion defined
\sect{Sec:dynamics}. Additionally, this baseline configuration also results in a
figure-of-eight pattern for the trailing edge. When the deviation is in
phase with the heaving motion, but with a reduced frequency twice as fast, the
pivot point yields the same pattern as the one described by
\cite{fry_aerodynamics_2003} and \cite{bos_influence_2008}. Increasing the
deviation amplitude widens this pattern and modifies the trajectory of the
trailing edge, which necessarily impacts how the foils perform the clap phase
when the minimum spacing $d^*$ is reached. Next, \fig{deviation_phase_shift}
shows the effect of modifying the phase shift $\psi$ for a configuration of
moderate deviation amplitude $l_0=0.035$. As seen in the figure, a positive
phase shift tilts the figure-of-eight pattern of the pivot point to the right
while a negative phase shift tilts it to the left. Moreover, when the phase
shift is equal to $\pm90\degree$ the figure of eight disappears and the
trajectory becomes the same for the upstroke and the downstroke of the foil
while maintaining a certain tilt. Lastly, the configuration with $l^*$ = 0.035
and $\psi=-125\degree$ which yielded the best efficiency presents a
figure-of-eight pattern tilted to the left for the pivot point, and
interestingly, the trajectory of the trailing edge becomes essentially the same
for the upstroke and downstroke. This configuration, which includes deviation,
changes the evolution of the effective angle of attack of each foil but
preserves a consistent trajectory of the trailing edges during the clap phase
where each of them essentially moves on the same line when approaching and
departing from the minimum spacing $d^*$.

\begin{figure}[tb]
\centering
\includegraphics[width=\linewidth]{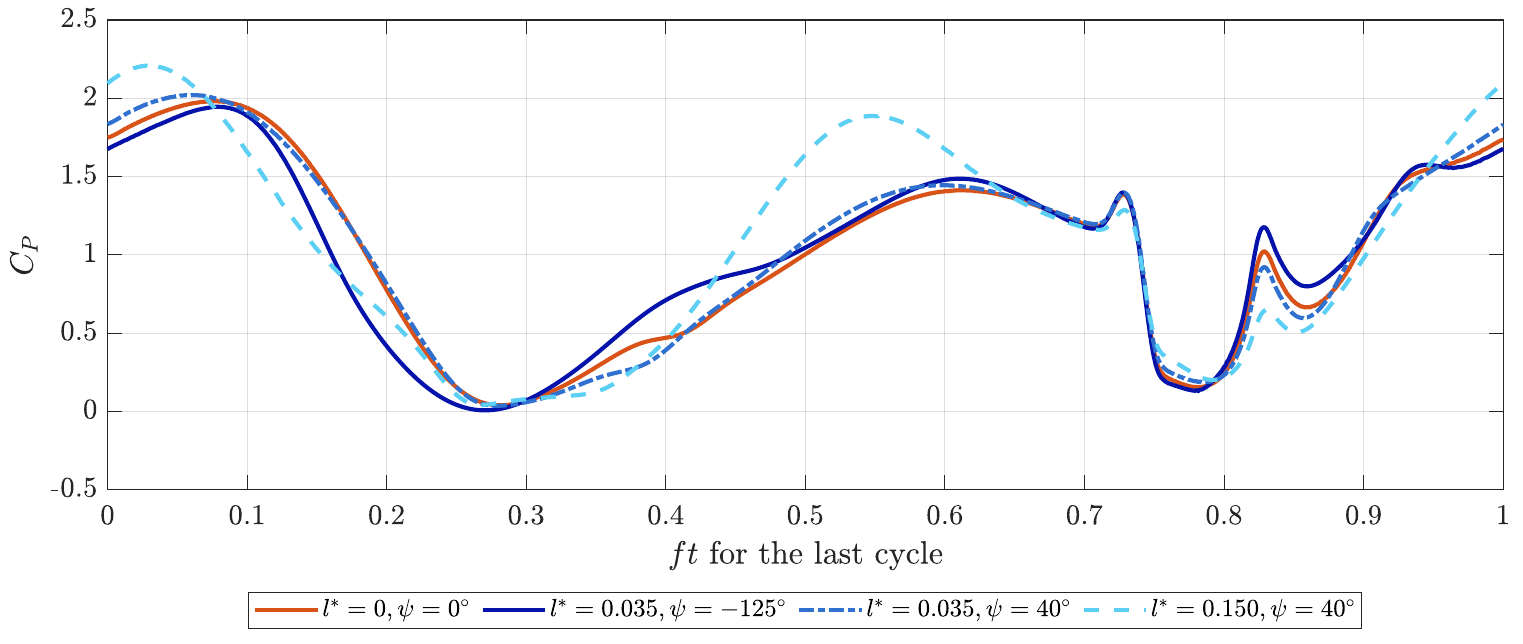} 
\caption{Comparison of the instantaneous power coefficient $C_P$ for different
combinations of $l^*$ and $\psi$.}
\label{fig_CP}
\end{figure}

\begin{figure}[b!]
\centering
\includegraphics[width=\linewidth]{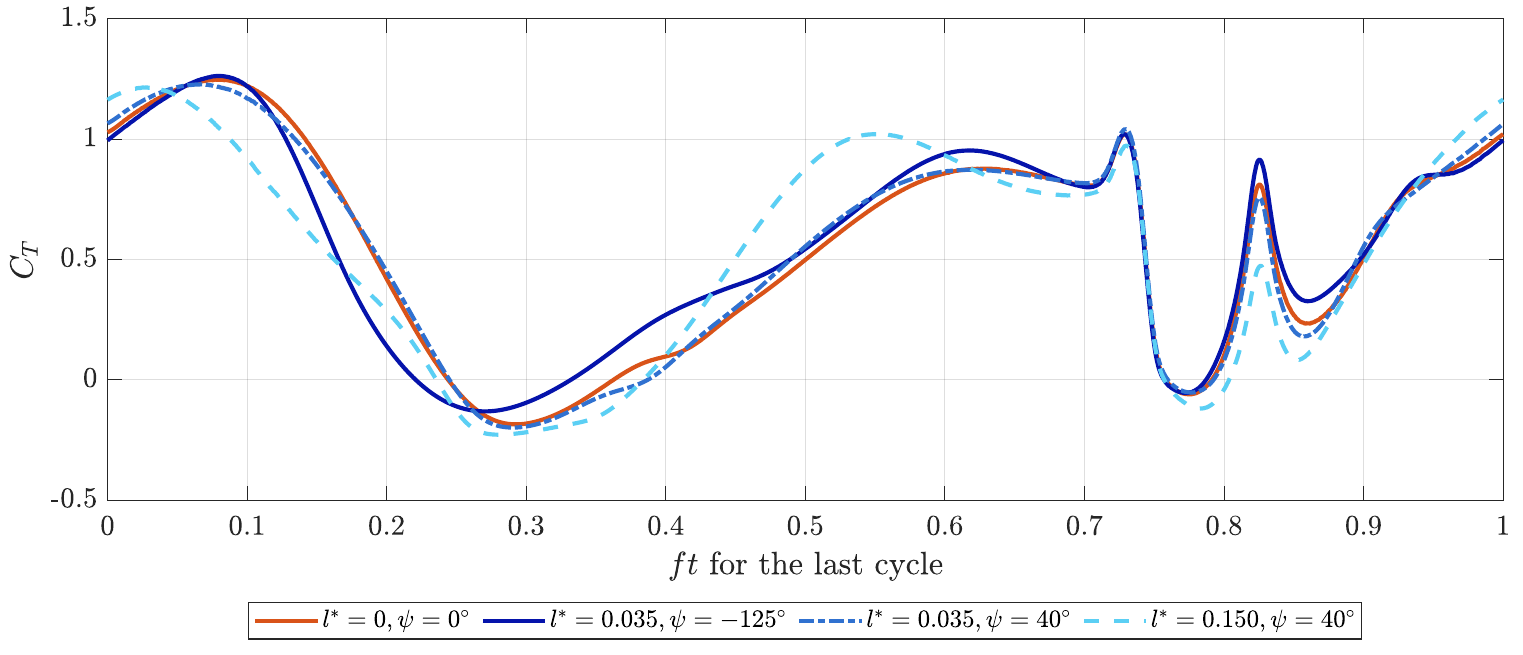} 
\caption{Comparison of the instantaneous thrust coefficient $C_T$ for different
combinations of $l^*$ and $\psi$.}
\label{fig_CT}
\end{figure}

\begin{figure}[htb]
\centering
\includegraphics[width=\linewidth]{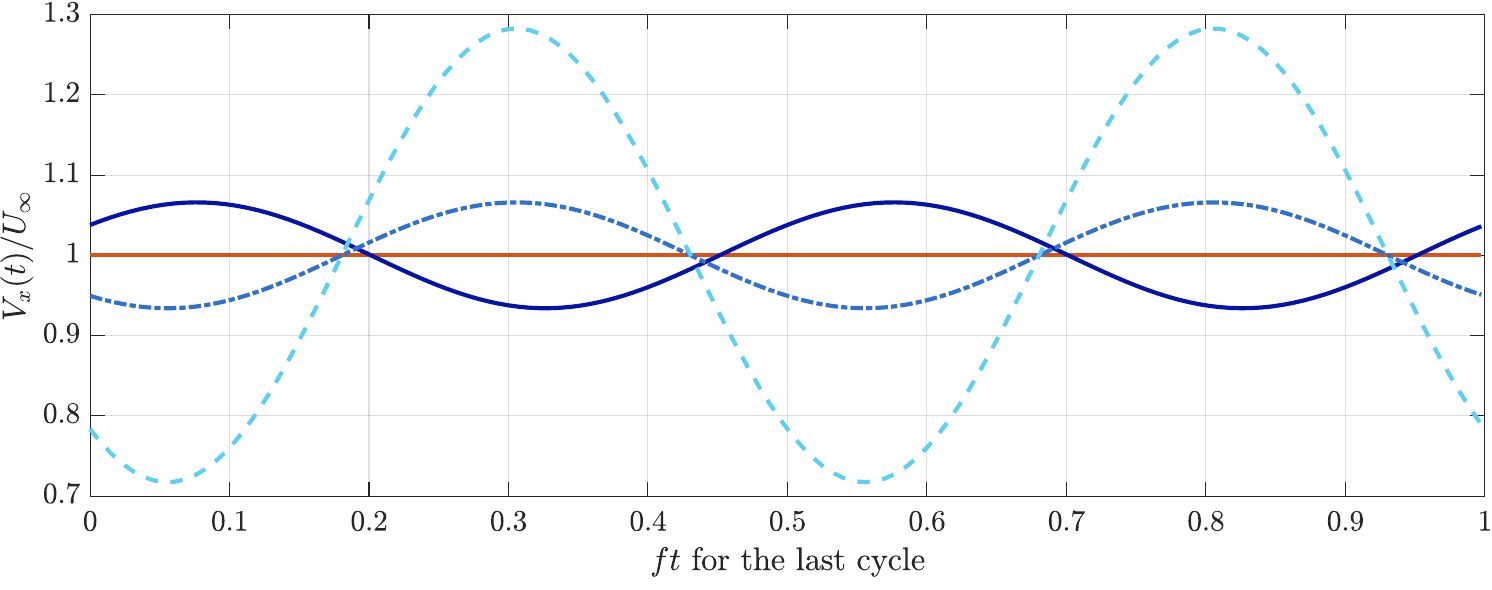} 
\includegraphics[width=\linewidth]{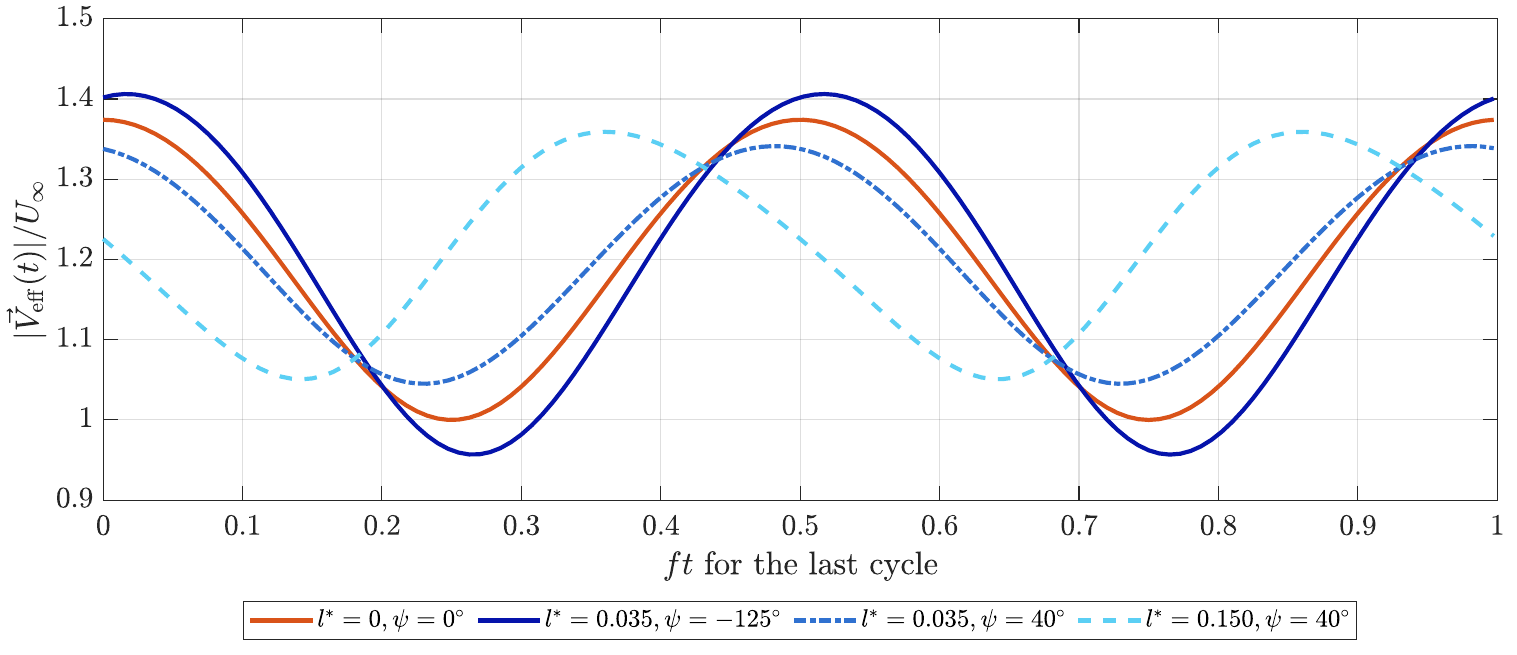} 
\caption{Horizontal component (top) and magnitude (bottom) of the effective
velocity $\vec{V}_{\text{eff}}$ for different combinations of $l^*$ and $\psi$.}
\label{effective_velocity}
\end{figure}

\begin{figure}[htb]
\centering
\includegraphics[width=\linewidth]{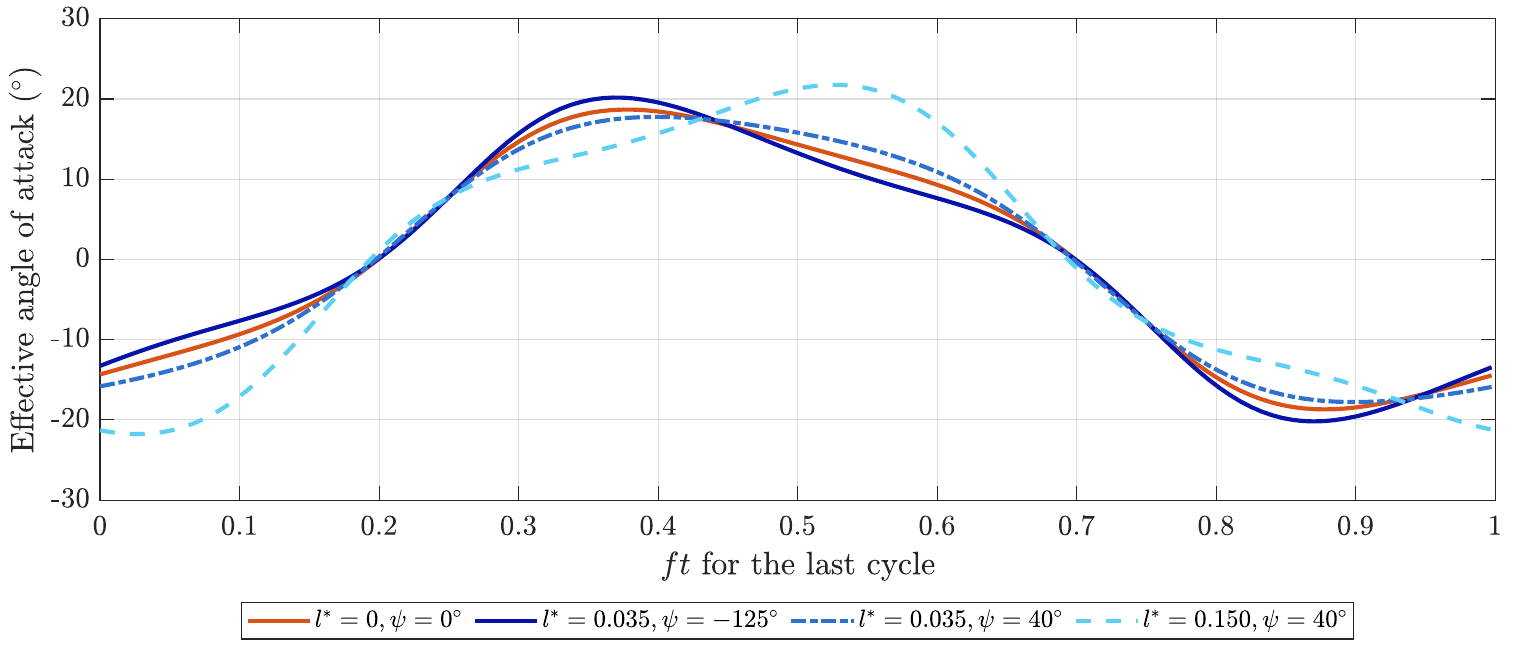} 
\caption{Effective angle of attack for different combinations of $l^*$ and $\psi$.}
\label{angle_of_attack}
\end{figure}

As briefly discussed, the different motion parameters change the figure-of-eight
pattern of the flapping foils thus impacting their effective velocity and angle
of attack, which can modulate the aerodynamic performances. \fig{fig_CP} and
\fig{fig_CT} present the evolution of instantaneous power and thrust
coefficients of cases with different parameters $l^*$ and $\psi$. The maximum
power and thrust coefficients are obtained near the middle of the upstroke close
to the beginning of the cycle, as shown in \fig{fig_CP} and \fig{fig_CT}. In
cases with no and moderate deviation, the peaks are similar and occur around the
instant $ft=0.08$. For the case with substantial deviation ($l^*$ = 0.150$,
\psi=40\degree$), the peaks are less important and occur earlier in the cycle.
This is due to the effective velocity which is out of phase in this
configuration compared to the other cases as observed in
\fig{effective_velocity}. This figure presents both the horizontal component of
the effective velocity and the magnitude of the effective velocity which
includes the vertical component related to the heaving motion of the foils (see
\eq{eq_effective_velocity}). Compared with the best case with deviation ($l^*$ =
0.035$, \psi=-125\degree$), the two other reported cases with deviation exhibit
effective horizontal velocity evolutions that are out of phase by $165\degree$,
thus almost perfectly reversing where the peaks and the troughs occur during one
cycle.  However, since the vertical component of the effective velocity is the
same in all cases and is higher in magnitude than the horizontal one, the
magnitude of the effective velocity is not as affected relative to the case
without deviation for cases with a moderate amount as seen in
\fig{effective_velocity}.  Indeed, the moderate normalized deviation amplitude
$l^*=0.035$ is a small fraction of the normalized heaving amplitude $h^*=1$ and
thus does not alter as much the distribution of the effective velocity as the
case with higher deviation with the normalized amplitude $l^*=0.150$. This phase
shift is also apparent in \fig{angle_of_attack} which presents the distribution
of the effective angle of attack of the different cases. The curves of the
effective angle of attack with no and moderate deviation are essentially the
same, while the one with substantial deviation has a larger amplitude that
occurs with a significant delay. This delay in the evolution of the effective
angle of attack thus explains why the first and second power and thrust peaks
observed during the upstroke and downstroke are respectively shifted. The peaks
are consequently of lower magnitude due to the effective velocity which is also
lower at this instant as seen in \fig{effective_velocity}.

As explained previously, the first peak in the power and thrust coefficient
curves is caused by the foils' acceleration after the stroke reversal occurring
during the clap phase of the previous motion cycle, imparting an important
impulse to the fluid which increases thrust \citep{sun_unsteady_2002}.
Afterward, the second stroke reversal of the foils occurs around $ft=0.30$ and
the foils thus begin the downstroke of the motion.  Another peak related to the
foils' acceleration is observed around $ft=0.615$ during the downstroke, similar
to one produced during the upstroke. 

The last two peaks observed during the cycle are directly related to the
clap-and-fling phenomenon. The one observed near $ft=0.729$ represents the
instant where the leading edges of the foils reach proximity due to the choice
of proper kinematics. As explained in \sect{Sec:results}, the configuration with
a phase shift between the heaving and pitching motion of $\phi = 105\degree$
generates a large positive pressure zone between the foils immediately to the
right of their leading edges as they reach proximity. The resulting pressure
gradient between the zones upstream and downstream of the leading edges of the
foils thus produces an important local thrust peak.  \fig{fig_CT} shows that
this thrust peak is not strongly affected by the different configurations where
deviation is present. At this instant, the effective angles of attack of all
cases are practically identical, while the effective velocities present
important differences, as seen in \fig{effective_velocity} and
\fig{angle_of_attack}. Thus, it can be concluded that changing the effective
velocity does not strongly influence the thrust and power coefficient curves. 

The last thrust peak which results directly from the clap-and-fling mechanism,
as explained in \sect{Sec:results}, occurs at $ft=0.826$.  It is observed that
the maximum value of the baseline case reported on \fig{fig_CT} (${C_T} = 0.78$)
is affected by the deviation component of the motion. Indeed, the optimal case
with moderate deviation ($l^*$ = 0.035$, \psi=-125\degree$) increases this
thrust peak to ${C_T} = 0.915$. At this instant, this case has a lower effective
velocity than the baseline case without deviation, as seen on
\fig{effective_velocity}. However, it was previously mentioned that the
effective velocity has little impact on performance.  On the other hand, the
effective angles of attack are different at this instant. Indeed, the
configuration without deviation presents a value of $-16.93\degree$, while the
optimal case with moderate deviation yields one of $-18.37\degree$. This
increase, slightly higher than $1\degree$, of the effective angle of attack thus
significantly modulates the instantaneous performance metrics, especially the
thrust coefficient. The lowest value of effective angle of attack at $ft=0.826$
is for the case with the most important tested deviation amplitude ($l^*$ =
0.150$, \psi=40\degree$). This low angle of attack ($\alpha=-12.32\degree$)
significantly decreases the thrust coefficient to ${C_T} = 0.47$.  Thus, a
proper phase shift between heaving and deviation $\psi$ can accentuate the
beneficial unsteady effects yielded by the clap-and-fling mechanism through the
modulation of the effective angle of attack. For example, the thrust peaks
obtained when the foils are close to each other during the clap phase can be
increased substantially. It was also shown that the deviation amplitude $l^*$
also greatly impacts the effective velocity as well as the effective angle of
attack of the foils during a cycle of motion, which is beneficial for moderate
values and impairs the flow for greater values such as $l^* = 0.150$.

\section{Conclusion}
\label{Sec:conclusion}

This study investigated oscillating foils employing the clap-and-fling mechanism
to generate thrust. Two-dimensional numerical simulations were conducted at a
Reynolds number of 800, with plate-shaped foils exhibiting sinusoidal pitching,
heaving, and deviation motions. Different phase shifts between these motions
were tested in the simulations. Two parametric studies, without and with
deviation, were performed at a reduced frequency of 0.15, ensuring periodic and
highly efficient results. In this regime, rapid convection of leading-edge
vortices in the wake prevents strong interaction with the two foils. The main
objectives of this study were first to characterize efficiency improvements
obtained through the clap-and-fling mechanism and to determine if the thrust
could also be increased. Second, the study aimed to determine if implementing
the deviation motion in a system subjected to the clap-and-fling mechanism with
optimized kinematics could further increase both thrust and efficiency. 

In cases without deviation, significant thrust and efficiency improvements were
obtained by reducing the minimum spacing between the foils and adjusting the
phase shift between the heaving and pitching motions. The best case produced a
0.542 efficiency and a thrust coefficient of 0.549. This occurred with a phase
shift of $105\degree$ and with a minimum spacing such that the two foils almost
touch at the trailing edge. This high-efficiency configuration represents a
significant improvement when compared to the best case of a single oscillating
foil which provided an efficiency of 0.478 at a phase shift of
$\phi=105\degree$ and a thrust coefficient of 0.36 at $\phi=110\degree$. In the
best case previously reported, this minimum spacing occurred at the trailing
edge during the motion of the foils, and it produced a low-pressure zone to
their upstream side due to the entrapment of the fluid between the two foils,
which resulted in a significant peak of thrust during the cycle at this instant.
Similarly, a high-pressure zone was observed between the foils downstream of the
leading edges as they come close to each other, which increases thrust and
efficiency. However, it was shown that this ideal scenario, where the pressure
gradient is favorable to the thrust production at a low minimum spacing, is
very sensitive to the phase shift. For example, increasing the phase shift to
$110\degree$ caused the minimum spacing to occur at the leading edge instead of
the trailing edge, decreasing both the thrust and the efficiency.

Further enhancements were achieved by including the deviation motion and
adjusting its amplitude and phase shift. In the most favorable configuration, a
remarkable 0.560 efficiency and an averaged thrust coefficient of 0.563 were
obtained.  This optimal configuration, with a phase shift of $\psi =
-125\degree$ and a moderate deviation amplitude of $l^* = 0.035$, produced
increases of 17.14\% in efficiency and 56.39\% in thrust coefficient compared to
the best single-foil case.  Implementing the deviation motion thus accentuates
the beneficial effects of the minimum spacing by favorably modulating the
effective angle of attack experienced by the foils near the thrust peak.
Moreover, the reported propulsive standard deviation was $\sigma_{C_T} = 0.412$.
This represents a slight decrease compared to the base case without deviation
($\sigma_{C_T} = 0.429$) indicating less transient force variations during the
motion.  However, increasing thrust solely through deviation comes at the cost
of having a slightly greater force variations throughout the motion.

Even though the wing kinematics studied in this work does not directly represent
the three-dimensional wing motions observed in flying species, it illustrates
that a simpler in-plane flapping motion exhibits similar benefits when combined
with deviation and clap-and-fling kinematics.  By using high-aspect-ratio wings
that would mitigate 3D effects, the kinematics described in this paper could
thus be used as a viable mechanical system to produce thrust. Nevertheless, the
exploration of three-dimensional effects could be undertaken through comparable
yet more extensive computational fluid dynamics simulations. This would allow
their influence on the clap-and-fling phenomenon to be quantified. These effects
are relevant to the current in-plane kinematics and to the more intricate
three-dimensional motions observed in nature.

\bibliographystyle{plainnat} 
\bibliography{Article_deviation_clap_fling}
 
\end{document}